\documentstyle[aps,epsfig,preprint]{revtex} 
\def\be{\begin{equation}} 
\def\ee{\end{equation}} 
\def\ba{\begin{eqnarray}} 
\def\ea{\end{eqnarray}} 
\def\no{\nonumber \\} 
\def\om{\omega} 
\def\gm{\gamma} 
\def\al{\alpha} 
 
\def\go{\Gamma_{1}} 
\def\gt{\Gamma_{2}} 
\def\gth{\Gamma_{3}}

\begin{document} 
\begin{titlepage} 
\pagestyle{empty} 
\vspace{1.0in} 
\begin{flushright} 
August 1999 
\end{flushright} 
\vspace{.1cm} 
\begin{center} 
\begin{large}  
{\bf Phases of Brans-Dicke Cosmology(II):Matter from NS-NS  sector} 
\end{large} 
\vskip 0.5in 
Sunggeun Lee\footnote{sglee@hepth.hanyang.ac.kr}, Chanyong Park 
\footnote{chanyong@hepth.hanyang.ac.kr} and 
Sang-Jin Sin\footnote{sjs@hepth.hanyang.ac.kr} \\ 
\vskip 0.2in 
{\small {\it 
Department of physics, Hanyang university, seoul, Korea}} 
\end{center} 
\vspace{1 cm} 
 
\begin{abstract} 
 
We study the cosmology of the Brans-Dicke(BD) theory coupled to perfect  
fluid type matter.  
In our previous works, the case where matter is coming from 
the Ramond-Ramond sector of the string theory was studied. 
Here we study the case where matter is coming from NS-NS sector. 
Exact solutions  are found and 
the cosmology is classified according  
to the values of $\gm$,the parameter of the equation of state and $\om$,
the BD parameter.  
We find that, in string frame, there are solutions without singularity
for some ranges of $\gm$ and $\om$.  
In Einstein frame, however, 
all solutions are singular.  
 
\end{abstract} 
\end{titlepage} 
 
\section{Introduction}  
The string theory is believed to be the most promising 
candidate to quantum gravity. So it is natural to expect  
that it will resolve some problems inherent in the general relativity 
like the initial singularity problem.  
In fact, in the regime  of Planck length curvature,  
quantum fluctuation is very large 
so that string coupling becomes large and consequently 
the fundamental string degrees of freedom 
are not weakly coupled {\it good} ones. 
Instead, solitonic degrees of freedom like solitonic p-branes\cite{duff} or 
D-brane\cite{pol} are more important. Therefore it is an interesting question 
to ask whether including these degrees 
of freedom resolve the initial singularity. 
 
The new gravity theory that can deal with such new degree of freedom 
should be a deformation of standard general relativity so that in a 
certain limit it should be reduced to the standard Einstein theory.   
The Brans-Dicke theory\cite{BD} is a generic deformation of  
the general relativity  
allowing variable gravity coupling.  
Therefore whatever is the motivation to modify 
the Einstein theory, the Brans-Dicke theory is the first one to be considered. 
As an example, low energy limit of the string theory 
contains the Brans-Dicke theory with a fine tuned deformation parameter 
($\omega=-1$) and it is extensively studied under the name of the string  
cosmology\cite{ve,g,sc}. 
Without knowing the exact theory of the p-brane cosmology, 
the best guess is that it should be a Brans-Dicke theory with matters. 
In fact there is some evidence for this \cite{duff},  
where it is found that the 
natural metric that couples to the p-brane is the  Einstein metric 
multiplied by certain power of dilaton field.  In terms of this new 
metric, the action that gives the p-brane solution  becomes 
Brans-Dicke action with definite deformation  parameter $\omega$ 
depending on p.  
 
In our previous works \cite{sung,park}, we  
studied the gas of  solitonic  p-brane \cite{duff} treated as 
a perfect fluid type matter in a Brans-Dicke theory  
allowing the equation of state parameter $\gamma$ arbitrary. 
We had studied the case where the perfect fluid does not couple to the  
dilaton like the matter in the Ramond-Ramond sector of the string theory. 
Here we study the opposite case where matter couples  
to the dilaton like those coming from NS-NS sector(see reference
\cite{muk}
for similar study). 
Exact solutions  are found both in string and Einstein frame and  
the cosmology is classified according  
to the values of $\gm$ and $\om$.  
In string frame we will find non-singular solutions 
for some ranges of $\gm$ and $\om$.  
In Einstein frame, however,   
we will find that all solutions are singular, unlike the string frame.  
 
The rest of this paper is organized as follows. 
In section II, we construct the action for the case when the matter 
coupled to the dilaton and find analytic solutions for the equations 
motion. 
In section III, the relation between  cosmic time $t$ and parameter 
$\tau$ is studied.  
In section IV, we consider the behavior of the scale factor $a$ as a 
function of $\tau$. 
Using these results, in section V, we 
study the scale factor as a function of the cosmic time $t$. 
In section VI, we study the asymptotic behavior of $a(t)$ and classify them
according to acceleration and deceleration phases.
Up to the section VI, all analyses were done in string frame. 
In section VII, we investigate the cosmology in Einstein frame. 
In section VIII, we summarize and conclude with some discussion.
 
\section{Action with solitonic NS-NS matter and its solutions} 
 
We begin with the four dimensional Brans-Dicke like string action  
of which matter is coupled to the dilaton field.  
\be 
S =\int d^4x \sqrt{-g} e^{-\phi} \left[ R-\omega(\nabla\phi)^2 + L_{m} \right]\ee 
Notice that the action differs from that in reference \cite{sung,park}
by the coupling of the matter Lagrangian $L_m$ with the dilaton factor.
By varying the action, we get equations of motion: 
\ba 
R_{\mu\nu} - \frac{1}{2}g_{\mu\nu}R &=& T_{\mu\nu} +  
\omega \{ {\nabla}_{\mu}\phi 
{\nabla}_{\nu}\phi -\frac{1}{2}g_{\mu\nu}(\nabla\phi)^2 \}  \no 
&+& \{ -\nabla_{\mu}\nabla_{\nu}\phi +\nabla_{\mu}\phi \nabla_{\nu}\phi 
+g_{\mu\nu} {\nabla}^2\phi -g_{\mu\nu} (\nabla\phi)^2 \}  \no 
R-2\omega{\nabla}^2 \phi + \omega (\nabla\phi)^2 + L_{m} &=& 0 . 
\ea 
Let's choose the metric as 
$$ds^2 = - N dt^2 + e^{2\alpha(t)} dx_i dx^i \;\;(i=1,2,3),$$   
where $N$ is a lapse function. 
We consider perfect fluid type matter whose energy-momentum 
tensor is   
$$T_{\mu\nu}= pg_{\mu\nu} +(p+\rho)U_{\mu}U_{\nu},$$  
satisfying the conservation law  
$$\dot\rho + 3 (p + \rho) \dot\alpha =0.$$ 
Using the equation of state, $p=\gm \rho,$ we get  
$$\rho = \rho_{0} e^{-3(1+\gamma)\alpha}.$$ 
So, we can rewrite the action as 
\be 
S = \int dt e^{3\alpha -\phi} \left[ \frac{1}{\sqrt{N}} \{-6{\dot\alpha}^2 + 
6{\dot\alpha}{\dot\phi} +  
\omega {\dot\phi}^2 \}-\sqrt{N} \rho_{0} e^{-3(1+\gamma)\alpha} 
\right]. 
\ee 
Now we define new variable $\tau$ by   
\be 
d\tau e^{3\alpha-\phi} = dt . \label{a}
\ee
Then, the action becomes 
\ba 
S &=& \int d\tau [ \frac{1}{\sqrt{N}} \{{-6\alpha^{\prime}}^2 + 
6\alpha^{\prime} \phi^{\prime} + 
\omega {\phi^{\prime}}^2 \} - 
\sqrt{N} \rho_{0} e^{3(1-\gamma)\alpha - 2\phi} ], \no 
&=& \int d\tau [\go {Y^\prime}^2+\gth {X^\prime}^2 -\rho_{0}e^{-2X}], 
\ea 
where  
\ba 
\go &=& -6+9(1-\gm) + \frac{9}{4}(1-\gm)^2 \om = \frac{9}{4}(1-\gm)^2 (\om
-\om_{\go}), \no 
\gt &=& 6+3(1-\gm)\om = 3(1-\gm)(\om -\om_{\gt}), \no 
\gth &=& \om - \frac{\gt^2}{4\go} = -\frac{3(2\om+3)}{\go}, \label{1} 
\ea
\ba
-2X &=& 3(1-\gm)\al - 2\phi, \no 
Y &=& \al + \frac{\gt}{2\go}X, \label{2} 
\ea
\ba
\om_{\go} &=& \frac{4(3\gm-1)}{3(1-\gm)^2}, \no 
\om_{\gt} &=& -\frac{2}{1-\gm}=\om_{\eta}.  
\ea 
 
From this action we get equations of motion: 
\ba 
Y^{\prime\prime} &=& 0, \no 
X^{\prime\prime} -\frac{\rho_{0}}{\gth}e^{-2X} &=& 0 . 
\ea 
The constraint equation is obtained by varying lapse function $N$: 
\be 
\go {Y^\prime}^2 + \gth {X^\prime}^2 +\rho_{0} e^{-2X} = 0. 
\ee 

The behavior of the solution depend crucially on the sign of $\go$. 
\begin{itemize} 
\item $\Gamma_1 < 0$ case:  
\ba 
X &=& \ln \left[ \frac{q}{c}\cosh{c\tau} \right], \no 
Y &=& A\tau + B . \label{b}
\ea 
where, $c,A,B$ and 
$q=\sqrt{\frac{\rho_0}{|\gth|}}$ are arbitrary real constants. 
Using the cosntraint equation, we can determine $A$
in terms of other variables 
\be 
A = c\sqrt{\frac{-\gth}{\go}} = c\frac{\sqrt{3(2\om+3)}}{|\go|}. 
\ee 
\item $\go > 0$ case: 
\ba 
X &=& \ln \left[ \frac{q}{c}|\sinh {c\tau}| \right], \no 
Y &=& A\tau + B . \label{c}
\ea  
\end{itemize} 
Having been found X and Y, $\al$ and $\phi$ can be found 
from the relation Eq.(\ref{2}). 

In next section, we will find $t(\tau)$ and 
find that its behavior depends on $\om$ and $\gm$.
\section{Phase space classification in terms of $t$ and $\tau$}  
\subsection{$\go < 0$ case}  
From Eqs.(\ref{a}) and (\ref{b}), $t(\tau)$ is found to be  
\be 
t-t_0 = \int d\tau e^{\frac{3}{2}(1+\gm)(A\tau +B)} 
\left[ \frac{q}{c}\cosh(c\tau) \right]^{-\frac{3\gt}{4\go}(1+\gm)-1} .
\ee 
It is easy to see $t(\tau)$ is a monotonic function. When $\tau$ goes to  
$\pm \infty$, $t$ can be approximately integrated to be 
\be 
t \sim \frac{1}{T_{\pm}}e^{T_{\pm}\tau} ,\label{13}
\ee 
where 
\be 
T_{\pm} =- \frac{3\sqrt{3(2\om+3)}}{2\go}(1+\gm) \mp  
\left[ \frac{3\gt}{4\go}(1+\gm) + 1 \right] . \label{14} 
\ee 
We have fixed $c=1$.
We will say that $t$ is supermonotonic function of $\tau$ when it is 
monotonic  
and $t$ runs the entire real line when $\tau$ does. When $t$ is  
supermonotonic function of $\tau$, the universe evolves from infinite past 
to infinite future. Otherwise, the universe has a starting(ending)
point at a finite cosmic time $t_i(t_f)$. The running range of $t$ depend
on the sign of $T$.
\ba 
-\infty < t < \infty \;\; &&{\rm if} \;\; T_{-} < 0 < T_{+}, \no 
-\infty < t < t_f \;\; &&{\rm if} \;\; T_{-} < 0 \;\; and \;\; T_{+} < 0, \no 
t_i < t < \infty  \;\; &&{\rm if} \;\; T_{-} > 0 \;\; and \;\; T_{+} > 0, \no 
t_i < t < t_f  \;\; &&{\rm if} \;\; T_{+} < 0 < T_{-}. 
\ea 

The solution for $T_- < 0$ is found to be 
\be 
-\frac{3}{2} < \om < -\frac{4}{3}, \;\; and \;\; \om > \om_\kappa,
\label{18}
\ee 
where we have defined 
$$\frac{(3\gm-5)}{3(1-\gm)} :=\om_\kappa.$$
To get Eq.(\ref{18}) we used following identity.
$$3(2\om +3) =(\frac{\gt}{2})^2 -\om \go. $$
The region II in Fig.1 is correspond to this solution.  
For $T_- > 0$, we have solution:
\be
\om > -\frac{3}{2}, \;\; \om < \om_\kappa, \;\; or \;\; \om > -\frac{4}{3} .
\ee
The region I and VII is satisfy this solution.

If $T_{+} < 0$, the solution to Eq.(\ref{14}) is
\be 
-\frac{3}{2} < \om < -\frac{4}{3}, \;\; and \;\; \om < \om_\kappa .
\ee 
The region I correspond to this solution.
For  $T_+ > 0$, we have 
\be
\om > -\frac{3}{2}, \;\; \om > \om_\kappa,\;\; or \;\; \om > -\frac{4}{3}.
\ee
The region II and VII satisfy this solution.
\subsection{ $\go > 0$ case}  
In this case the solution $X(\tau)$ in Eq.(\ref{c}) has a singularity 
at $\tau=0$. So we have to 
look at the behavior of $t$ near $\tau =0$ carefully. 
From Eqs.(\ref{a}) and (\ref{c}), the $t(\tau)$ can be written as  
\be 
t-t_0 = \int d\tau e^{\frac{3}{2}(1+\gm)[\frac{\sqrt{3(2\om+3)}}{\go} 
c\tau + B]}\left[ \frac{q}{c}|\sinh(c\tau)| \right]^{-\frac{3\gt}{4\go}(1+\gm)-1} .
\ee 
The asymptotic behavior of $t$ in the limit $\tau \rightarrow \pm \infty$, 
is given by
\be
t \sim \frac{1}{T_{\pm}}e^{T_{\pm}\tau},
\ee
where 
\be 
T_{\pm} =\frac{3\sqrt{3(2\om+3)}}{2\go}(1+\gm) \mp  
\left[ \frac{3\gt}{4\go}(1+\gm) + 1 \right] . 
\ee 
Notice the difference from Eq.(\ref{13}).

The condition $T_{-} < 0$ gives solution: 
\be 
\om > -\frac{3}{2},\;\; \om < \om_\kappa, \;\; and \;\; \om > -\frac{4}{3}. 
\ee 
There is no region satisfying this solution.
For $T_- > 0$ the solution is
\be
\om > -\frac{3}{2},\;\; \om > \om_\kappa, \;\; or \;\; \om < -\frac{4}{3}.  
\ee
Therefore III, IV, V and VI satisfies this solution.

The solution for $T_{+} < 0$ is 
\be 
\om > -\frac{3}{2}, \;\; \om > \om_\kappa, \;\;and \;\; \om > -\frac{4}{3}. 
\ee 
The region V and VI satisfy this solution.
For $T_+ > 0$ the solution is
\be
\om > -\frac{3}{2}, \;\; \om < \om_\kappa, \;\; or \;\; \om < -\frac{4}{3}.
\ee
The region III and IV satisfy this solution.

So far our analysis is parallel to the previous section. However,
we have to pay attention to the behavior of $t(\tau)$ near $\tau =0$.
In the limit $\tau \to 0$, 
\be 
t \sim \frac{ {\rm sign}(\tau)}{1-\eta} |\tau|^{1-\eta} 
\ee 
where $\eta = \frac{3\gt}{4\go}(1+\gm) +1$.  
Notice $t(\tau)$ is regular at $\tau = 0$, if $\eta < 1$. 
When $\eta > 1$, $t(\tau)$ is singular at $\tau =0$.  
So we consider $t(\tau)$ in the region $-\infty < \tau < 0$ and 
$0 < \tau < \infty$ separately. The condition  
$\eta > 1$ is equivalent to 
\be 
\om > \om_{\gt} . 
\ee 
When $\tau$ goes to zero from the below, 
\be 
t \sim \frac{(-\tau)^{1-\eta}}{\eta-1}, 
\ee 
which means $t \to \infty$ as $\tau \to -0$. 
On the other hand, when $\tau$ goes to zero from the above, 
\be 
t \sim -\frac{(-\tau)^{1-\eta}}{\eta-1}, 
\ee 
so that $t \to -\infty$ as $\tau \to +0$.  

From these analysis, we see that the parameter space of $\gm$ and
$\om$ is divided into seven regions as depicted in Fig.1.

\begin{figure}[hbt] 
\centerline{\epsfig{file=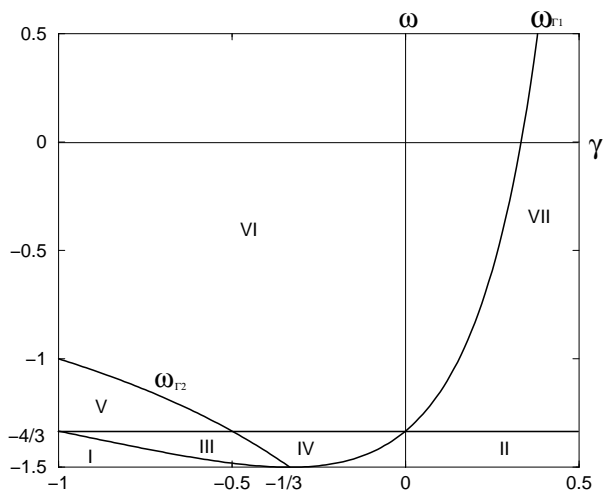,width=8cm}} 
\caption{\small The phases classified by the 
         relation between $t$ and $\tau$. } 
\end{figure} 

We summarize the results that are found in this section. 
\begin{itemize}
\item Region I: $T_{+} < 0$, $T_{-} > 0$, $\go < 0$; 
$t$ evolves from initial time $t_i$ to final 
time $t_f$ for $\tau \in (-\infty,\infty)$.
\item Region II: $T_{-} < 0$, $T_{+} > 0$, $\go < 0$; 
$t$ evolves 
from negative infinity to positive infinity for $\tau \in 
(-\infty, \infty)$. 
\item Region III: $T_{+} > 0$, $T_{-} > 0$, $\go > 0$;  
$t$ evolves from initial 
time $t_i$ to positive infinity for $\tau \in (-\infty,\infty)$.
\item Region IV: $T_{-} > 0$, $T_{+} > 0$, $\go > 0$; 
Since $\tau =0$ is singular, the region of $\tau$ is divided into two
regions. $t$ evolves from initial time $t_i$ to positive infinity 
for $\tau \in (-\infty, 0)$; 
$t$ evolves from negative infinity to positive infinity 
for $\tau \in (0, \infty)$.
\item Region V: $T_{+} < 0$, $T_{-} > 0$, $\go > 0$;  
$t$ evolves from initial 
time $t_i$ to final time $t_f$ for $\tau \in (-\infty,\infty)$. 
\item Region VI: $T_{+} < 0$, $T_{-} > 0$, $\go > 0$;
Since $t$ is singular at $\tau =0$, we should divide into two regions. 
$t$ evolves from initial  
time $t_i$ to positive infinity for $\tau \in (-\infty, 0)$ and  
negative infinity to final time $t_f$ for $\tau \in (0, \infty)$. 
\item Region VII: $T_{-} > 0$, $T_{+} > 0$, $\go < 0$;  
$t$ evolves from initial 
time $t_i$ to positive infinity for $\tau \in (-\infty, \infty)$.  
\end{itemize} 
\section{The behavior of the scale factor} 
We now study the phases of the cosmology by looking at the scale factor 
$a(\tau)=\exp(\alpha(\tau))$.
\subsection {$\go < 0$ case}
In this case $\alpha(\tau)$ in scale factor $e^{\al(\tau)}$ is given by 
\be 
\al{(\tau)} = \frac{c\sqrt{3(2\om+3)}}{-\go}\tau + B -\frac{\gt}{2\go} 
\left[ \ln{\frac{q}{c}\cosh(c\tau)} \right] .
\ee 
In the limit $\tau \to \pm \infty$, the scale factor can be written as 
\be 
a(\tau) \sim e^{H_{\pm}\tau} ,
\ee 
where the $H_{\pm}$ is defined by 
\be 
H_{\pm} = -\frac{c\sqrt{3(2\om+3)}}{\go} \mp \frac{\gt}{2\go}c . \label{29}
\ee 

Eq.(\ref{29}) for $H_{-} < 0$ gives the solution
\be 
\om > \om_{\gt}, \;\;and \;\; \om  < 0 . 
\ee 
The region II in Fig.2 satisfies this solution. 
For $H_- > 0$ the solution is
\be
\om < \om_{\gt}, \;\;or \;\; \om  > 0 . 
\ee
The region I and VI satisfy this solution.

If $H_{+} < 0$, the solution is given by 
\be 
\om < \om_{\gt}, \;\; and \;\; \om  < 0 .
\ee 
The region I satisfies this solution.
For $H_+ > 0$ the solution is
\be
\om > \om_{\gt}, \;\; or \;\; \om > 0 .
\ee
The region II and VI satisfy this solution. 
\subsection{$\go >0$ case}  
In this case, the $\al(\tau)$ in scale factor $e^{\al(\tau)}$ is given by  
\be 
\al{(\tau)} = \frac{c\sqrt{3(2\om+3)}}{\go}\tau + B -\frac{\gt}{2\go} 
\left[ \ln{\frac{q}{c}|\sinh(c\tau)|} \right] .
\ee 
In the limit $\tau \to \pm \infty$, $a(\tau)$ is given by 
\be 
a(\tau) \sim e^{H_{\pm}\tau},
\ee 
where $H_{\pm}$ is defined by
\be 
H_{\pm} = \frac{c\sqrt{3(2\om+3)}}{\go} \mp \frac{\gt}{2\go}c .
\ee 

The solution to the condition $H_{-} < 0$ is  
\be 
\om < \om_{\gt}, \;\;and \;\; \om > 0. 
\ee 
There is no region satisfying this solution.
For $H_- > 0$ the solution is
\be
\om > \om_{\gt}, \;\;or \;\; \om < 0. 
\ee 
The satisfying region is III, IV and V.

The solution for $H_{+} <0$ case is 
\be 
\om > \om_{\gt}, \;\; and \;\; \om > 0 .
\ee 
The region V satisfies this solution.
For $H_+ > 0$ the solution is given by
\be
\om < \om_{\gt}, \;\; or \;\; \om < 0 .
\ee
The region III and IV satisfy this solution.

Now we consider $\tau \to 0$ limit. In this limit $a(\tau)$ is
approximately given by 
\be 
a(\tau) \sim  |\tau|^{-\frac{3(1-\gm)(\om -\om_{\gt})}{2\go}}. 
\ee 
Therefore, if $\om > \om_{\gt}$, $a(\tau)$ goes to infinite as $\tau 
\to 0$. 
From these analyses we get a phase diagram Fig.2.   

\begin{figure}[hbt] 
\centerline{\epsfig{file=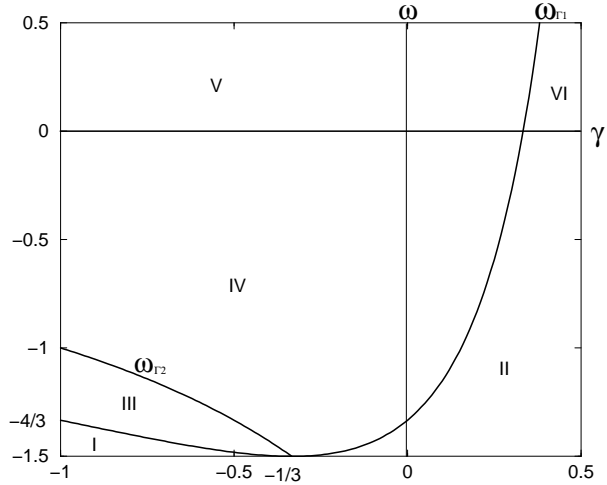,width=8cm}} 
\caption{\small The parameter space is classified by  
       the scale factor $a(\tau)$ } 
\end{figure} 

Summarizingly, we have following cases.
\begin{itemize}
\item Region I: $H_{-} > 0$, $H_{+} < 0$, $\go < 0$; The
scale 
factor $a(\tau)$ goes to zero size as $\tau \to \pm \infty$. 
\item Region II: $H_{-} < 0$, $H_{+} > 0$, $\go < 0$;
$a(\tau)$ goes 
to infinity as $\tau \to \pm \infty$. 
\item Region III: $H_{-} > 0$, $H_{+} > 0$, $\go > 0$;  
In this region, the behavior of $t$ is not singular, so we need not 
consider the behavior of $a(\tau)$ at $\tau =0$ where the scale 
factor vanishes. 
$a(\tau)$ goes to  
zero as $\tau \to -\infty$ and $a(\tau)$ goes to infinity as $\tau  
\to \infty$. 
\item Region IV: $H_{-} > 0$, $H_{+} > 0$, $\go > 0$; 
The $a(\tau)$ goes to 
infinite size as $\tau$ goes to zero for $\tau \in (-\infty, 0)$
and zero size as
$\tau \to -\infty$. $a(\tau)$ goes to infinity  as $\tau$ goes to zero 
for $\tau \in (0, \infty)$ 
and goes to positive infinity as $\tau \to \infty$. 
\item Region V: $H_{-} > 0$, $H_{+} < 0$ ,$\go > 0$; 
$a(\tau)$ goes to infinity  as $\tau$ goes to zero and $a(\tau)$ goes 
to zero as $\tau$ goes to negative infinity for $\tau \in (-\infty, 0)$.  
$a(\tau)$ goes to infinity as $\tau$goes to zero and $a$ goes tozero
$\tau$  
infinity for $\tau \in (0, \infty)$.   
\item Region VI: $H_{-} > 0$, $H_{+} > 0$, $\go < 0$; 
In the limit of  
$\tau \to -\infty$, $a(\tau)$ goes to zero size. In the  
limit of $\tau \to \infty$, $a(\tau)$ goes to infinite size. 
\end{itemize}
 
\section{Phases of the cosmology}     
In previous sections we have studied $t(\tau)$ and $a(\tau)$. 
From all considerations of these results, we can classify the 
parameter space of 
$\gm$ and $\om$ by the behavior of $a(t)$ into sixteen phases. In Fig.3, 
we show the phase diagram. 
In asymptotic region where $\tau \to \pm\infty$, we can write the scale  
factor $a(t)$ as: 
\ba 
a(t) \sim [T_{-}(t-t_i)]^{\frac{H_{-}}{T_{-}}} \no 
a(t) \sim [T_{+}(t-t_f)]^{\frac{H_{+}}{T_{+}}} . \label{39}
\ea  
From these relations we see that the behavior of the scale factor $a(t)$ 
depends on the sign of $T_{\pm}$ and the value of $H_{\pm}/T_{\pm}$ 
determins acceleration or deceleration of the scale factor
which is discussed below.
In Fig.4 - Fig.6, we show the behavior of scale factor by
numerical study.
The $-$ sign  in $V-$ indicates
the branch for $\tau \in  (-\infty, 0)$. Similarly, $V+$ means 
$\tau \in (0,\infty)$.
 
\begin{figure}[htb] 
\centerline{\epsfig{file=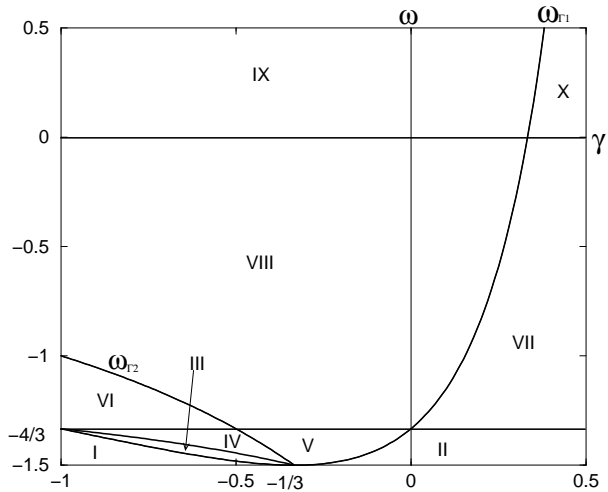,width=8cm}} 
\caption{\small The parameter space is clssified by the behavior 
        of $a(t)$ } 
\end{figure} 

\begin{itemize}
\item Region I, $T_{-} >0, T_{+} <0, H_{-} > 0$ and $H_{+} < 0$. The universe 
evolves from zero size at a finite initial time $t_i$ to a zero size 
at a finite final time $t_f$.  
This region contains the matter for inflation 
$\gm =-1$. However, we see in the limit $\tau \to -\infty$ the 
cosmic time $t$ approaches to $t_i$. 
\item Region II, $T_{-} < 0, T_{+} > 0, H_{-} < 0$ and $H_{+} > 0$. The 
universe evolves from infinite to infinite size as $t$ runs from  
negative infinity to positive infinity. 
This is the region where we can find the non-singular 
behavior of $a(t)$. 
\item Region III, $T_{+} > 0, T_{-} > 0, H_{-} > 0$ and $ H_{+} > 0$.
The universe 
evolves from zero size to infinite as $t$ runs from finite initial time  
$t_i$ to infinity for $\tau \in (-\infty ,\infty)$. During evolution
the universe goes to zero as 
$\tau$ goes to zero. 
So we divided into two branches at $\tau =0$.   
\item Region IV, $T_{-} > 0, T_{+} > 0, H_{-} > 0$ and $H_{+} > 0$. The  
universe evolves from zero to infinite size as $t$ runs from finite $t_i$ to
infinity. Like region $III$ we divided into two branches since the 
universe becomes zero as $\tau$ goes to zero. 
\end{itemize} 
By numerical study we depicted the behavior of scale 
factor $a(t)$.
 
\vspace{-3cm} 
\begin{figure} 
\unitlength 1mm 
\begin{center} 
\begin{picture}(70,100) 
\put(-40,10){\epsfig{file=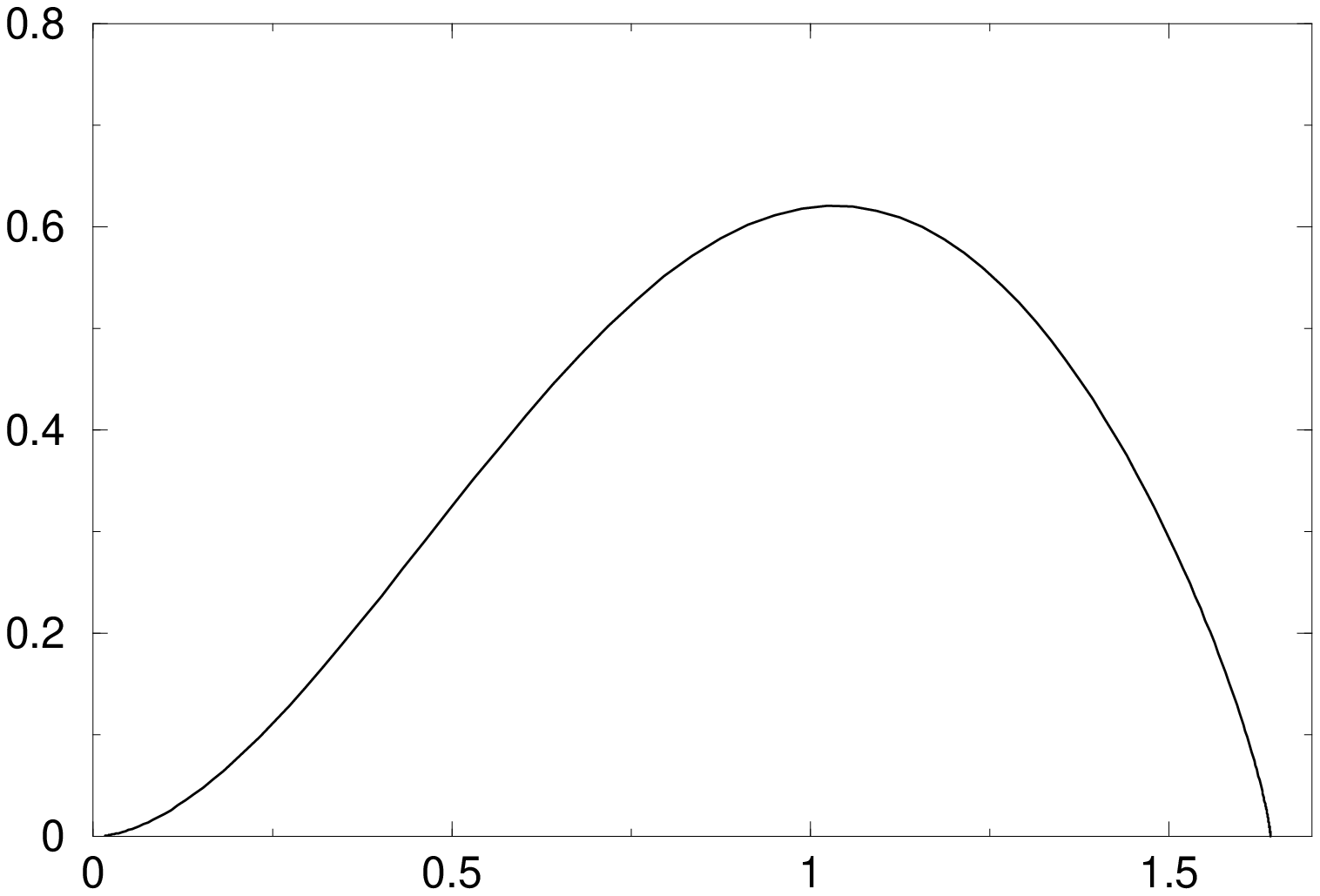,width=5cm,height=6cm}} 
\put(15,10){\epsfig{file=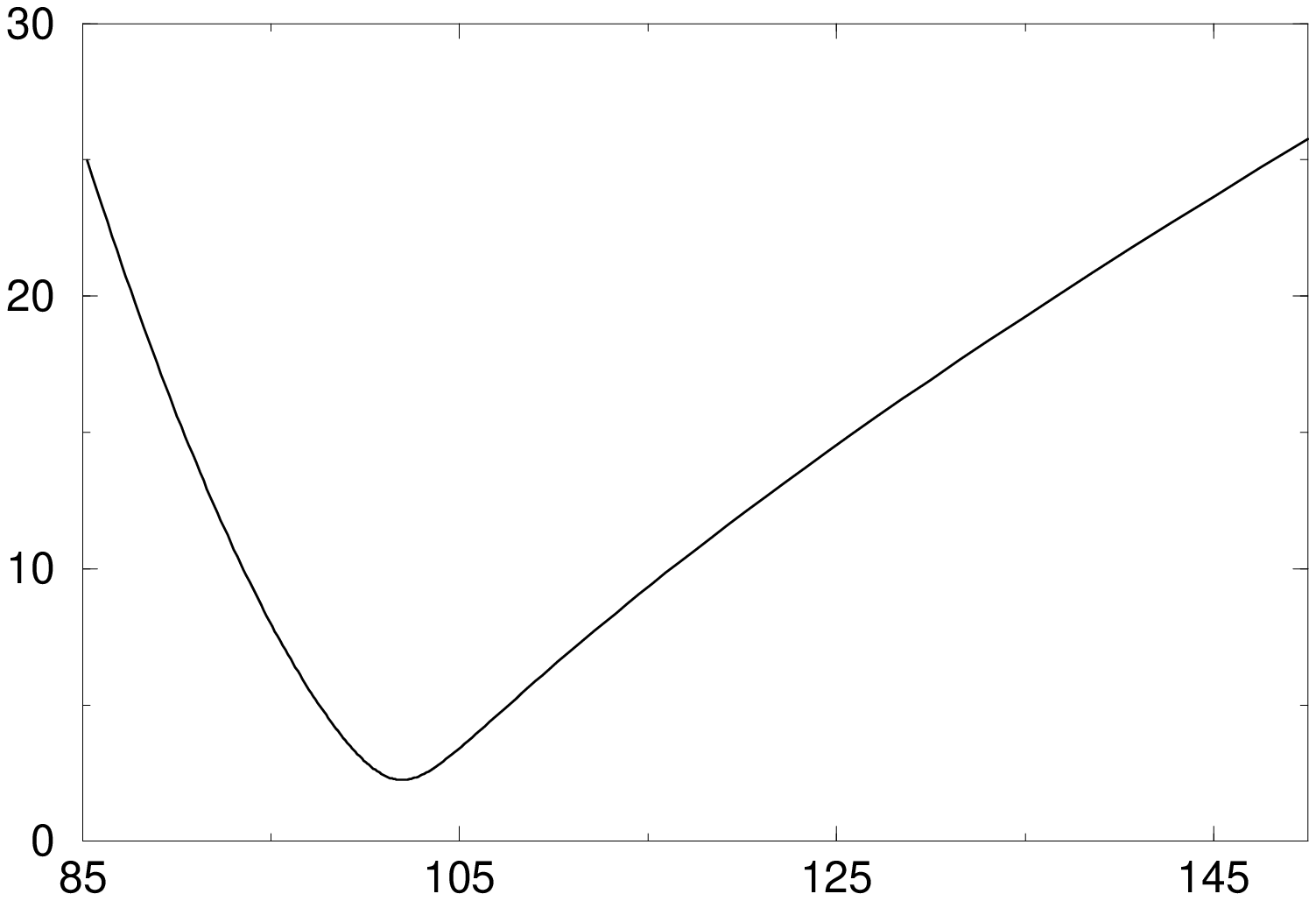,width=5cm,height=6cm}} 
\put(70,10){\epsfig{file=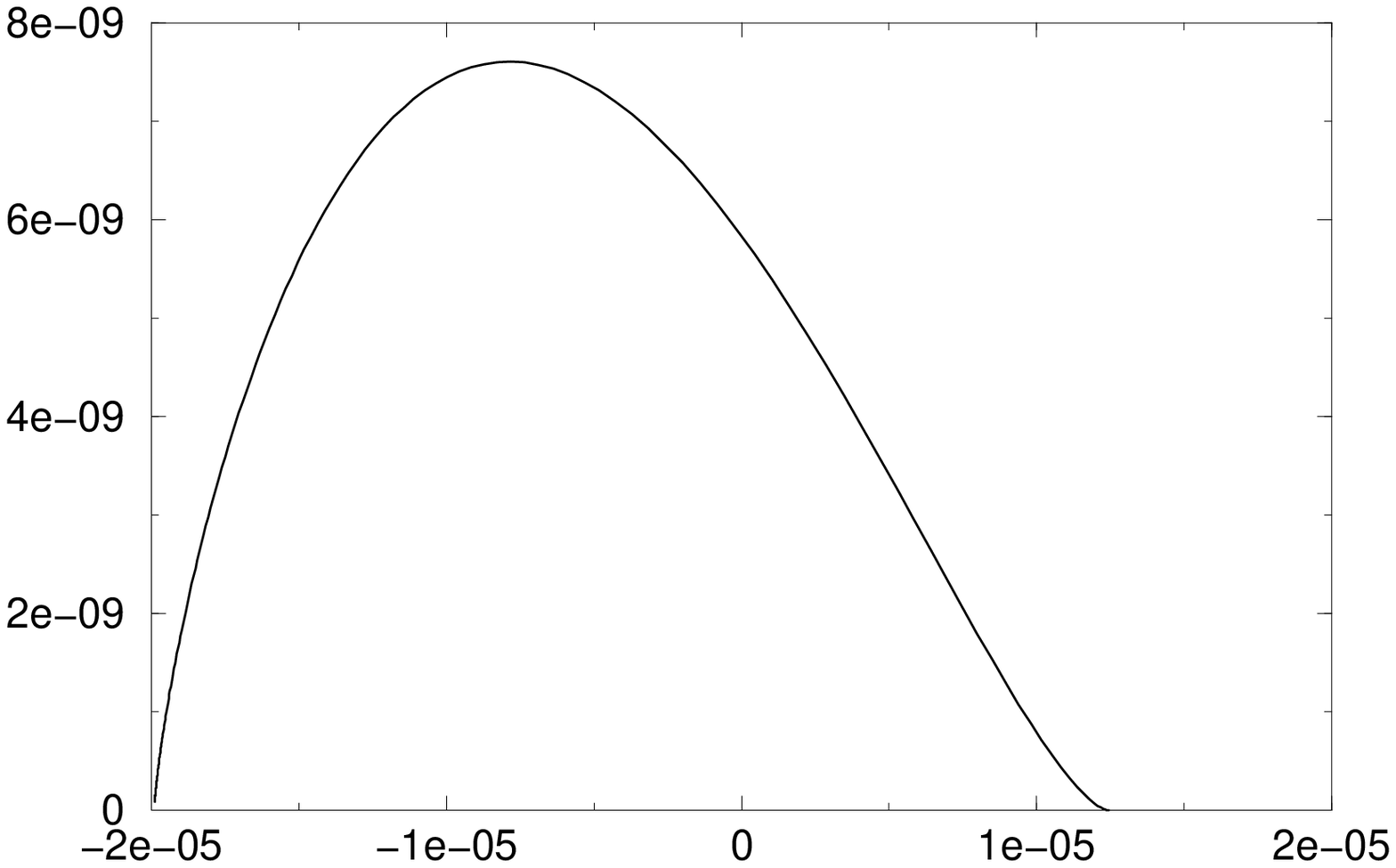,width=5cm,height=6cm}} 
\put(-40,-60){\epsfig{file=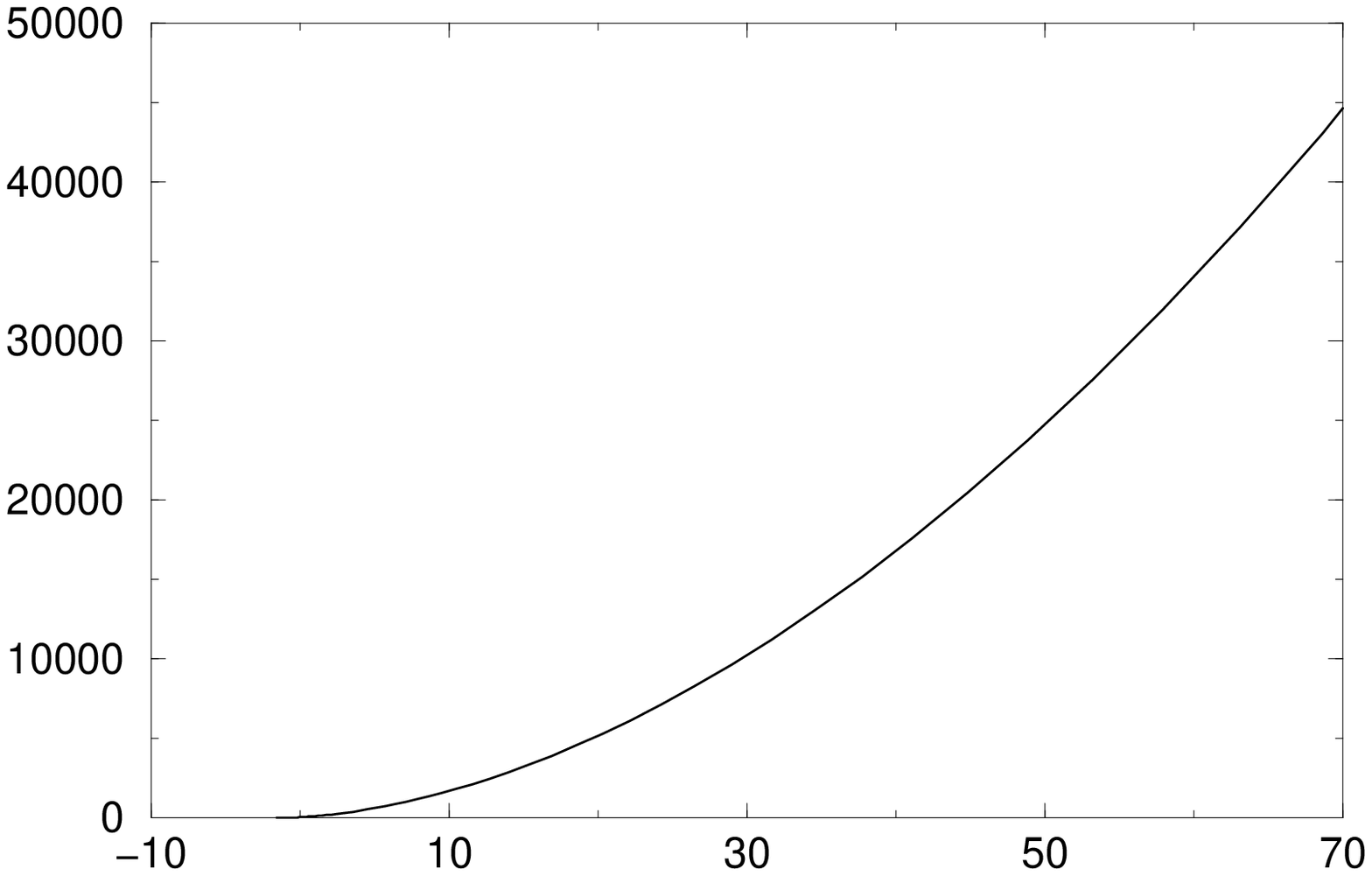,width=5cm,height=6cm}} 
\put(15,-60){\epsfig{file=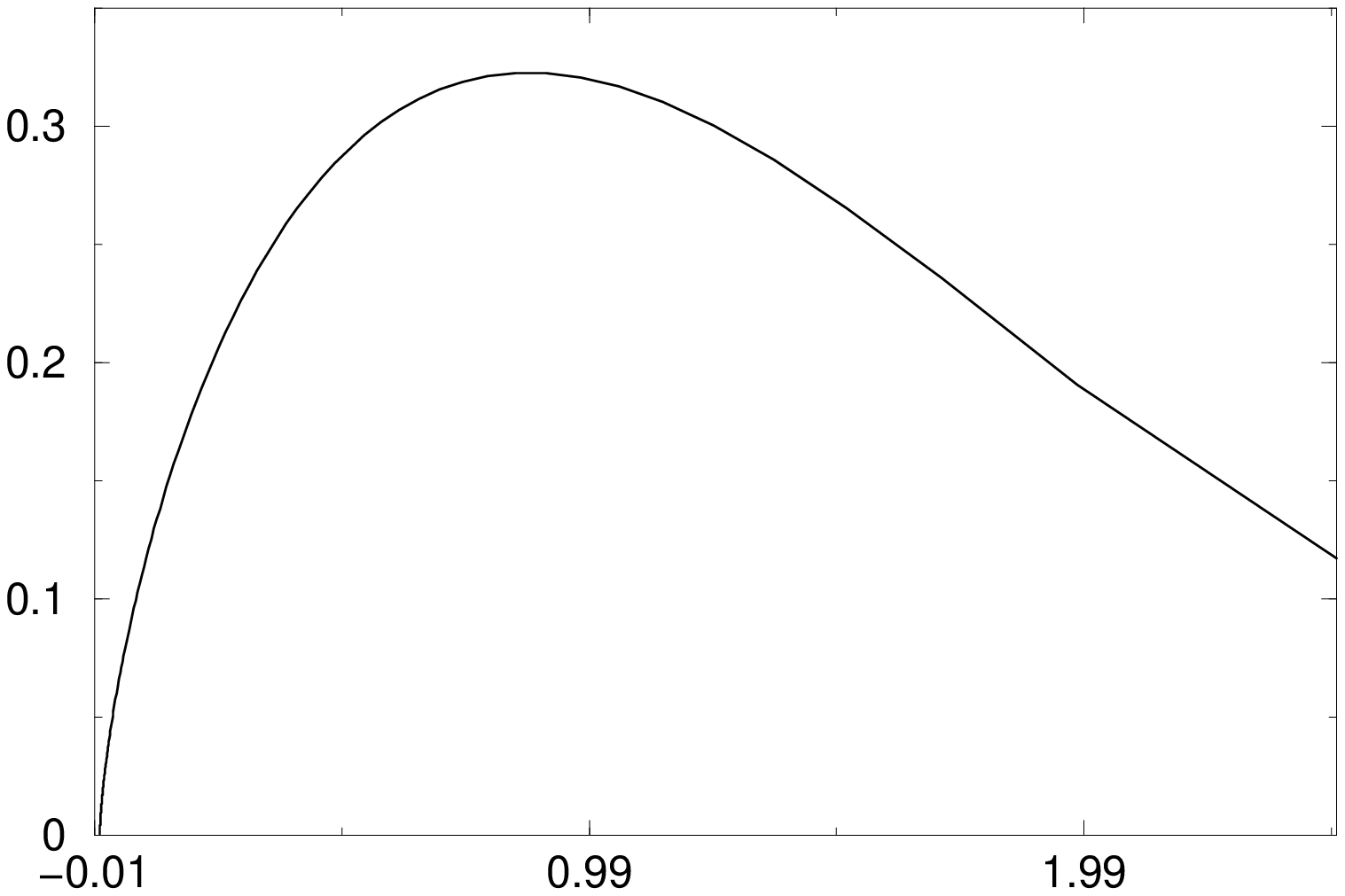,width=5cm,height=6cm}} 
\put(70,-60){\epsfig{file=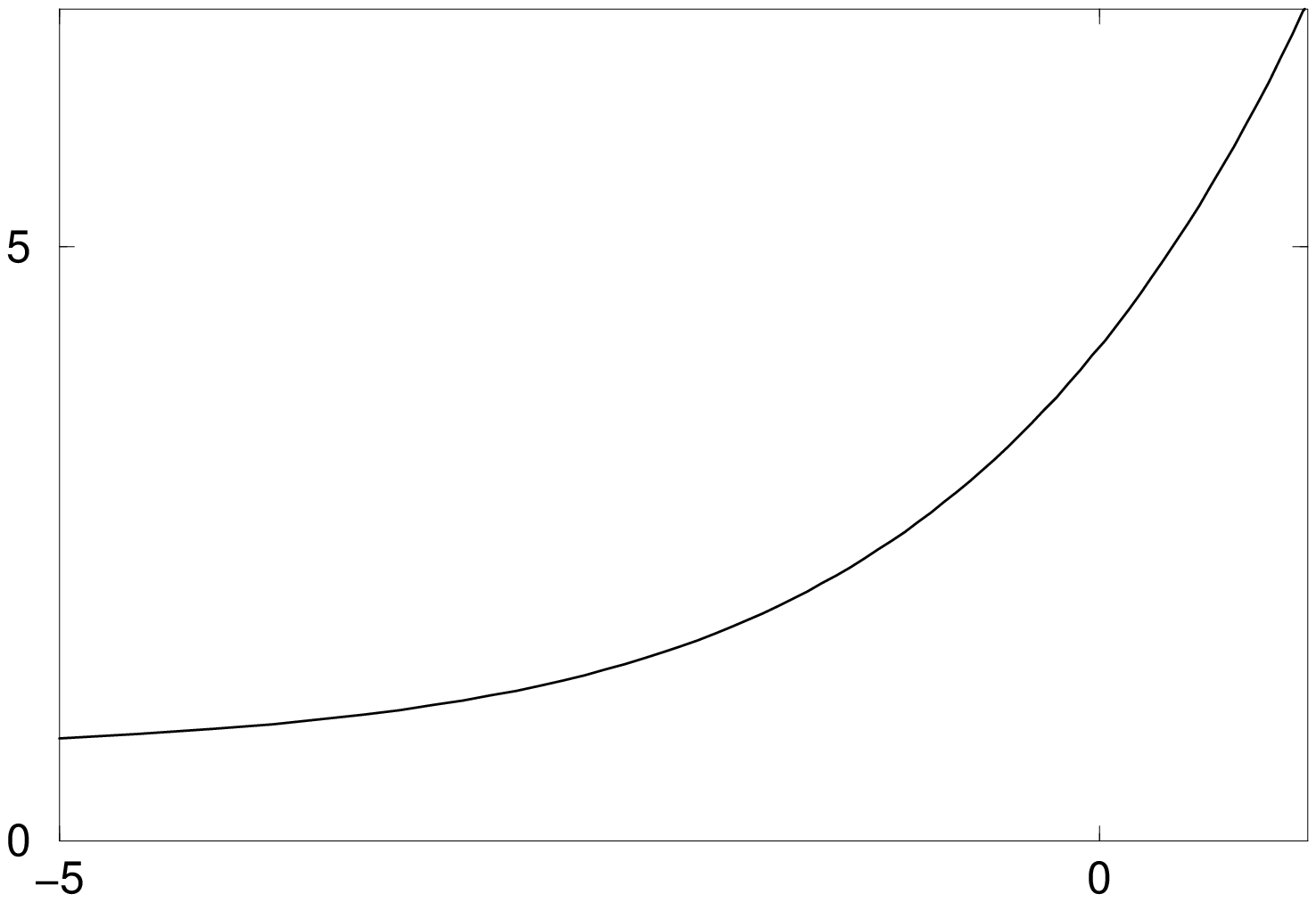,width=5cm,height=6cm}} 
\put(-35,5){$(a)$ phase $I$} 
\put(20,5){$(b)$ phase $II$} 
\put(75,5){$(c)$ phase $III-$} 
\put(-35,-65){$(d)$ phase $III+$} 
\put(20,-65){$(e)$ phase $IV-$} 
\put(75,-65){$(f)$ phase $IV+$} 
\end{picture} 
\end{center} 
\vspace{6cm} 
\caption{The behavior of the scale factor from phase $I$ to $IV$} 
\end{figure}

\begin{itemize} 
\item Region V, $ T_{-} > 0, T_{+} > 0, H_{-} > 0$ and $H_{+} > 0$.  
The universe evolves from zero to infinite size as $t$ runs finite initial  
time $t_i$ to infinity and the universe evolves from infinite to infinite 
as 
$t$ runs negative infinity to infinity.  
\item Region VI, $T_{-} > 0, T_{+} < 0, H_{-} > 0$ and $H_{+} > 0$. 
The universe evolves from zero to zero size as $t$ runs from initial time  
$t_i$ to infinity and the universe evolves from zero to infinite size as 
$t$ runs negative infinity to finite final time $t_f$. 
\item Region VII, $T_{-} > 0, T_{+} > 0, H_{-} < 0$ and $H_{+} > 0$.  
The universe evolves from infinite to infinite size as $t$ runs 
from finite initial time $t_i$ to infinity. 
\end{itemize}

\vspace{-3cm} 
\begin{figure} 
\unitlength 1mm 
\begin{center} 
\begin{picture}(70,100) 
\put(-40,10){\epsfig{file=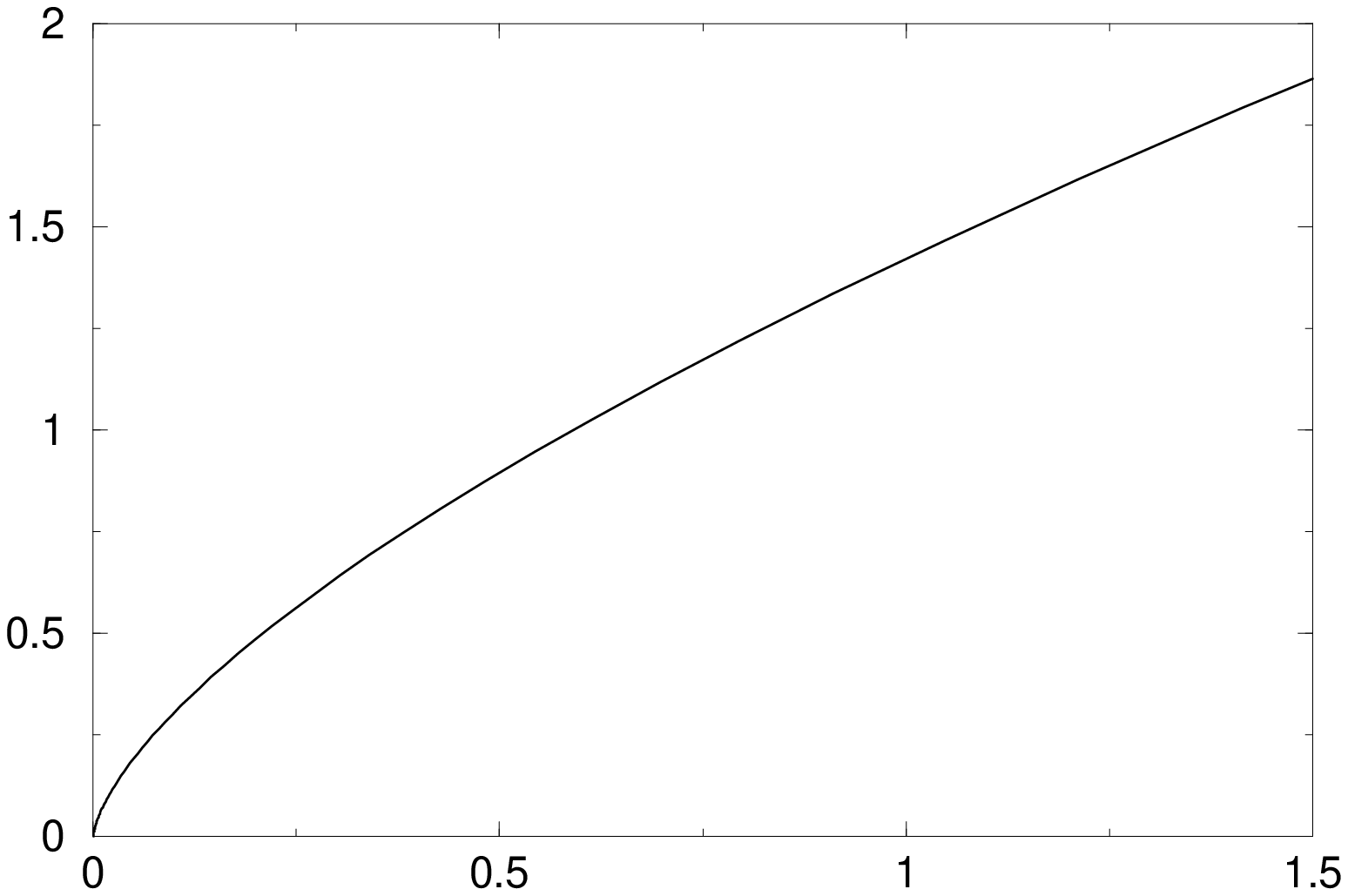,width=5cm,height=6cm}} 
\put(15,10){\epsfig{file=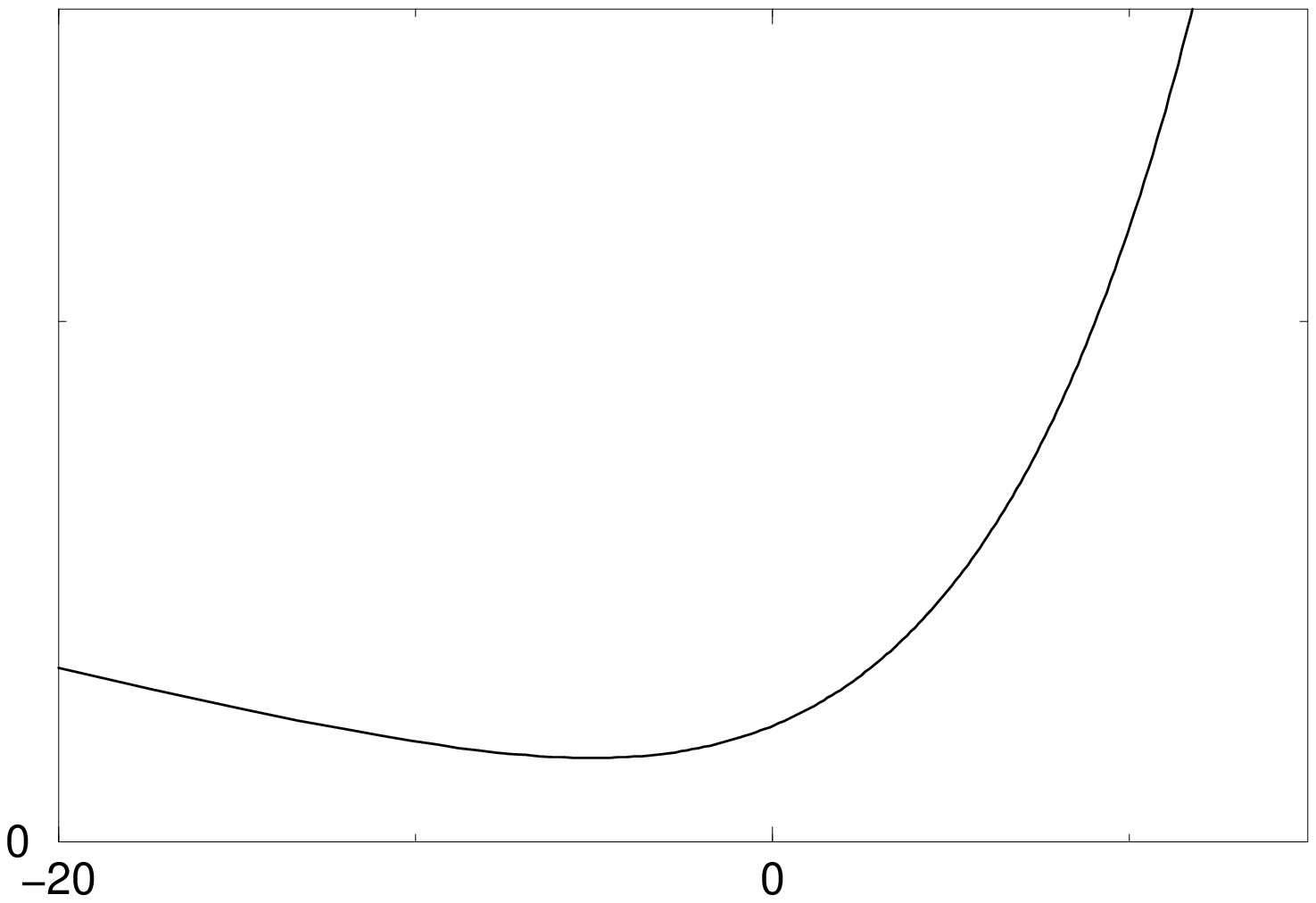,width=5cm,height=6cm}} 
\put(70,10){\epsfig{file=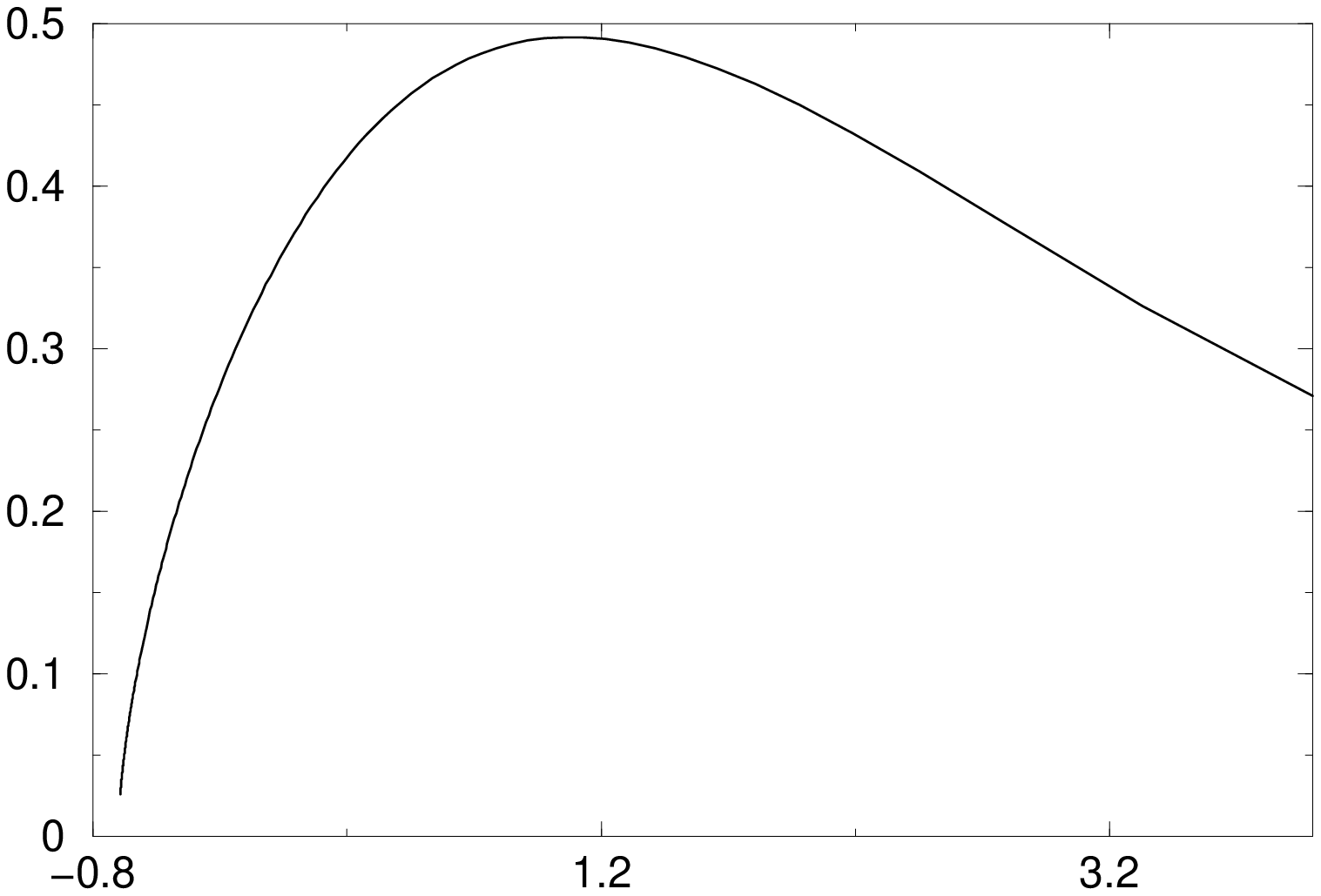,width=5cm,height=6cm}} 
\put(-40,-60){\epsfig{file=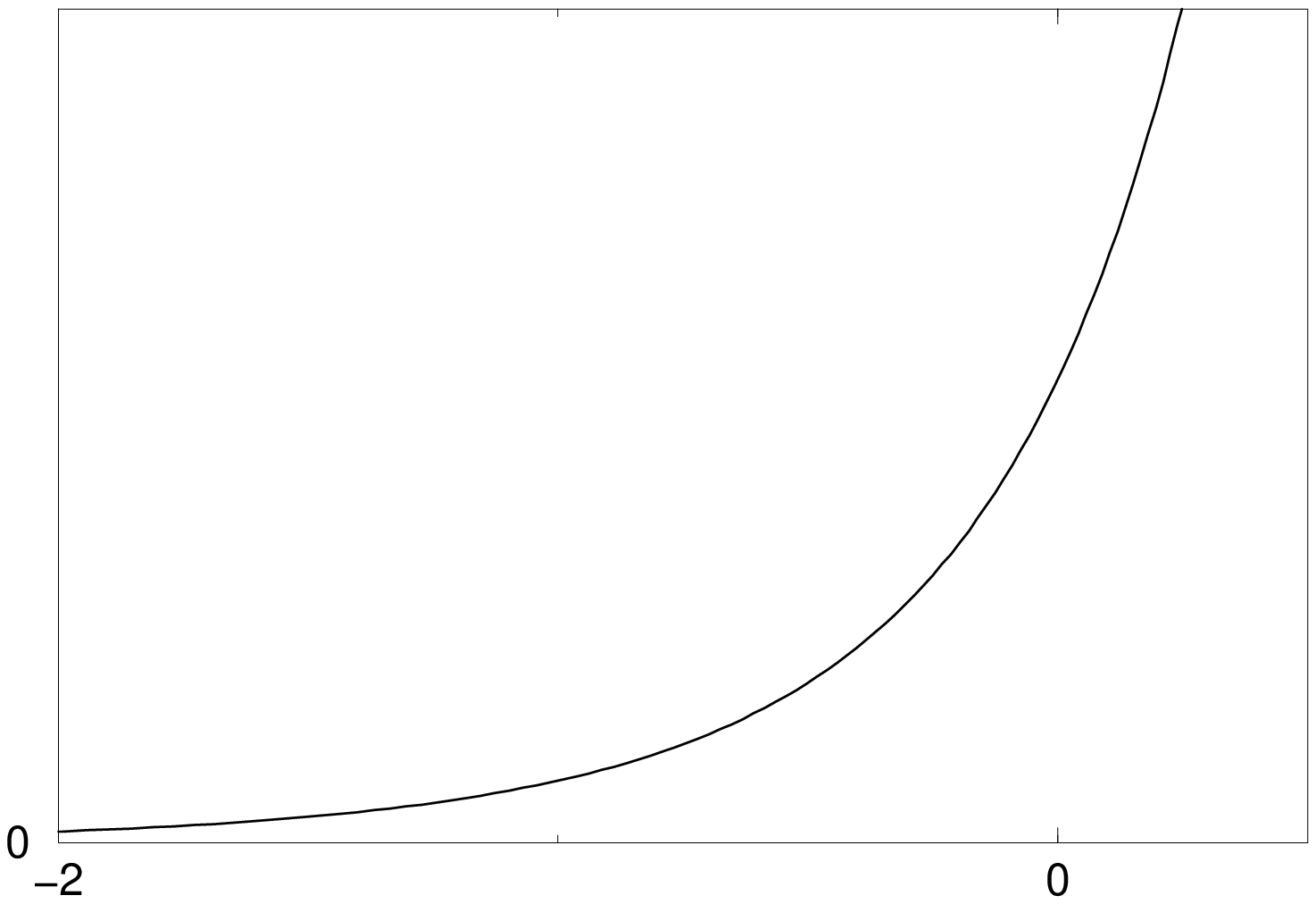,width=5cm,height=6cm}} 
\put(15,-60){\epsfig{file=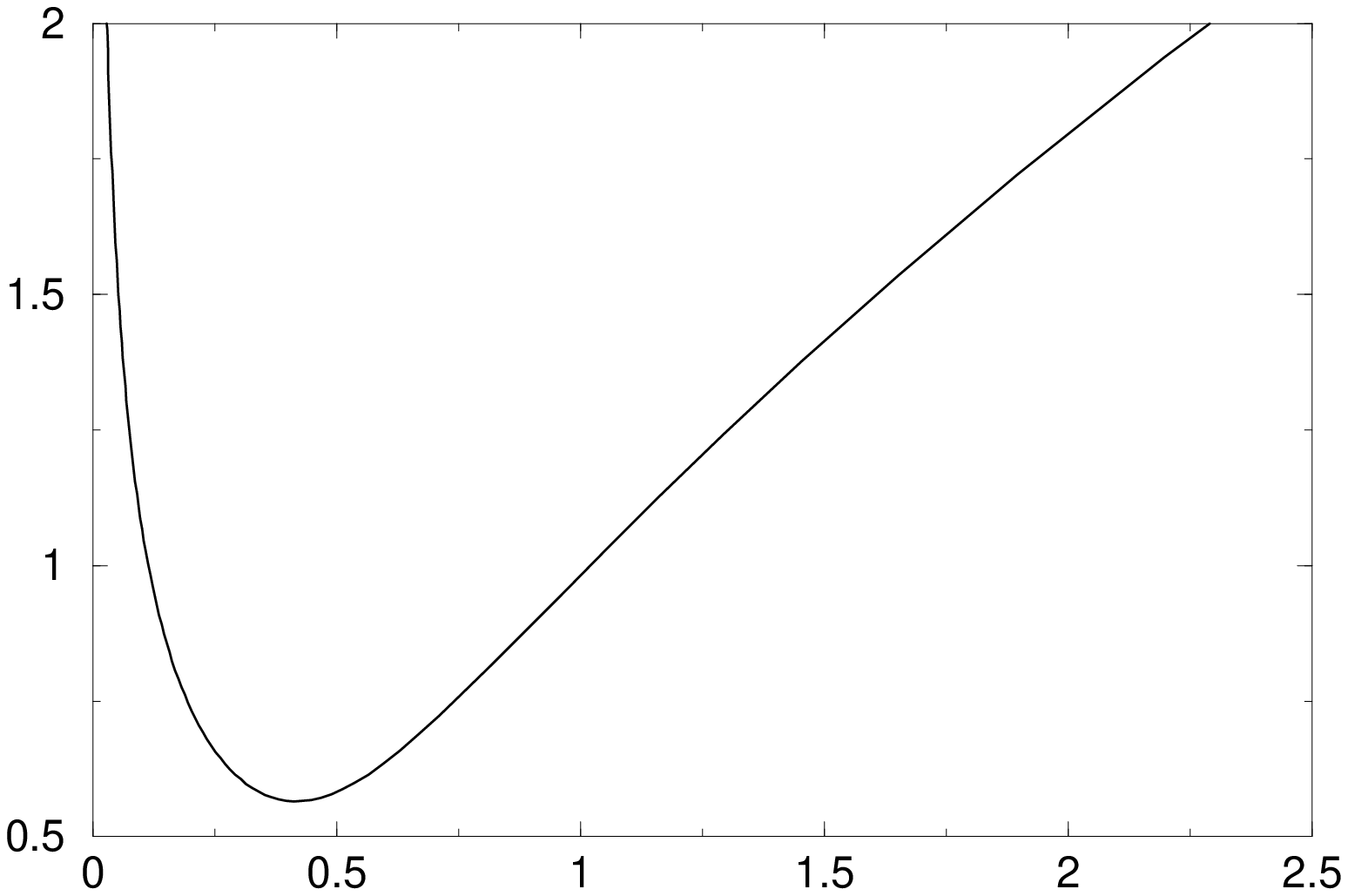,width=5cm,height=6cm}} 
\put(-35,5){$(a)$ phase $V-$} 
\put(20,5){$(b)$ phase $V+$} 
\put(75,5){$(c)$ phase $VI-$} 
\put(-35,-65){$(d)$ phase $VI+$} 
\put(20,-65){$(e)$ phase $VII$} 
\end{picture} 
\end{center} 
\vspace{6cm} 
\caption{The behavior of the scale factor from phase $V$ to $VII$} 
\end{figure} 
 
\begin{itemize}  
\item Region VIII, $T_{-} > 0, T_{+} < 0, H_{-} > 0$ and $H_{+} > 0$. The 
universe evolves from zero to infinite size as $t$ runs from finite  
initial time $t_i$ to infinity and the universe evolves from infinite to 
infinite size as $t$ runs negative infinity to finite final time $t_f$. 
\item Region IX, $T_{-} > 0, T_{+} < 0, H_{-} > 0$ and $H_{+} < 0$. The 
universe evolves from zero to infinite size as $t$ runs from finite  
initial time $t_i$ to infity and the universe evolves from infinite size 
to zero as $t$ runs from negative infinity to finite final time $t_f$. 
\item Region X, $T{-} > 0, T_{+} >0 , H_{-} > 0$ and $H_{+} > 0$.  
The universe evolves from zero to infinite size as $t$ runs from finite  
initial time $t_i$ to infinity. 
\end{itemize}  
 
\vspace{-3cm} 
\begin{figure} 
\unitlength 1mm 
\begin{center} 
\begin{picture}(70,100) 
\put(-40,10){\epsfig{file=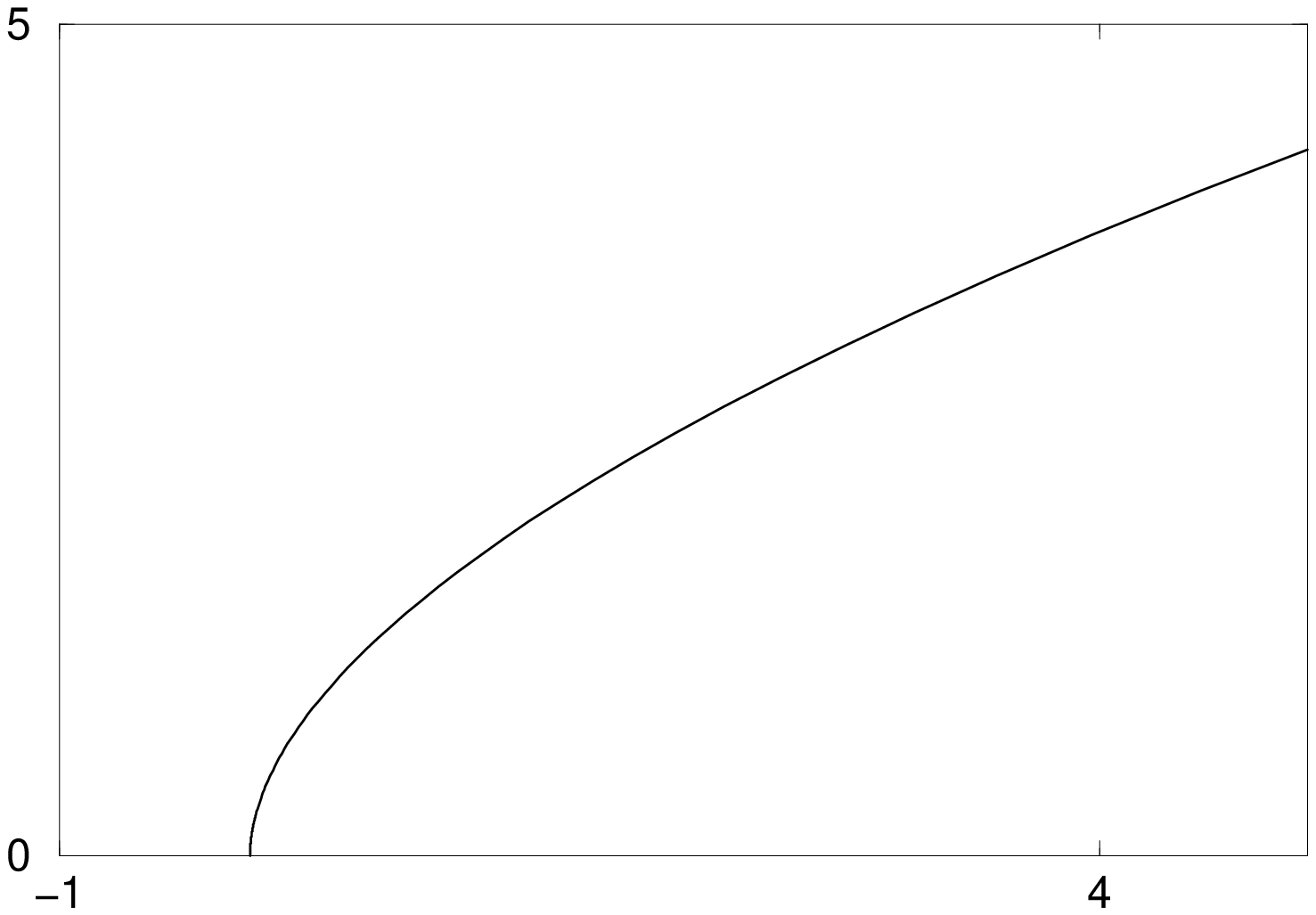,width=5cm,height=6cm}} 
\put(15,10){\epsfig{file=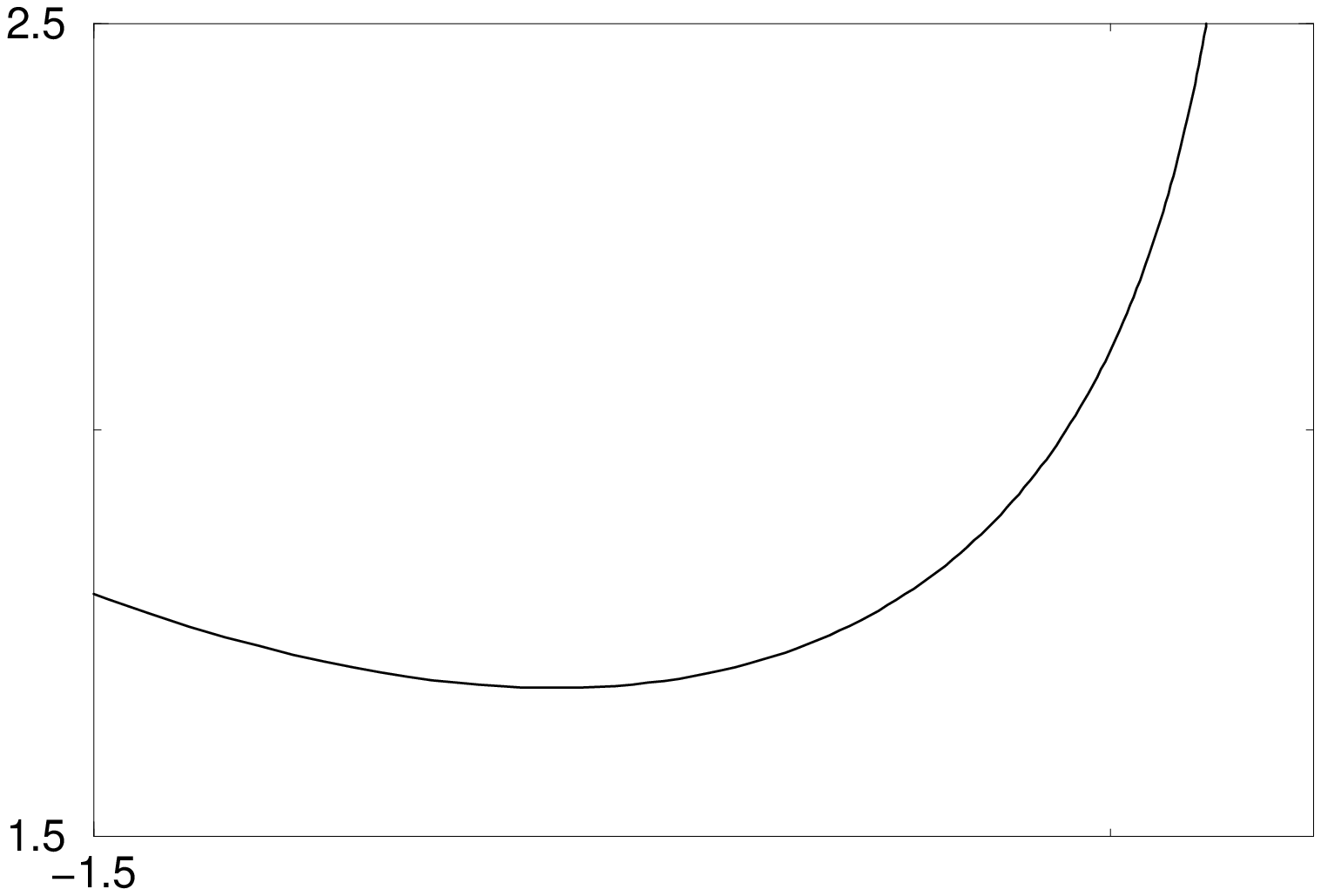,width=5cm,height=6cm}} 
\put(70,10){\epsfig{file=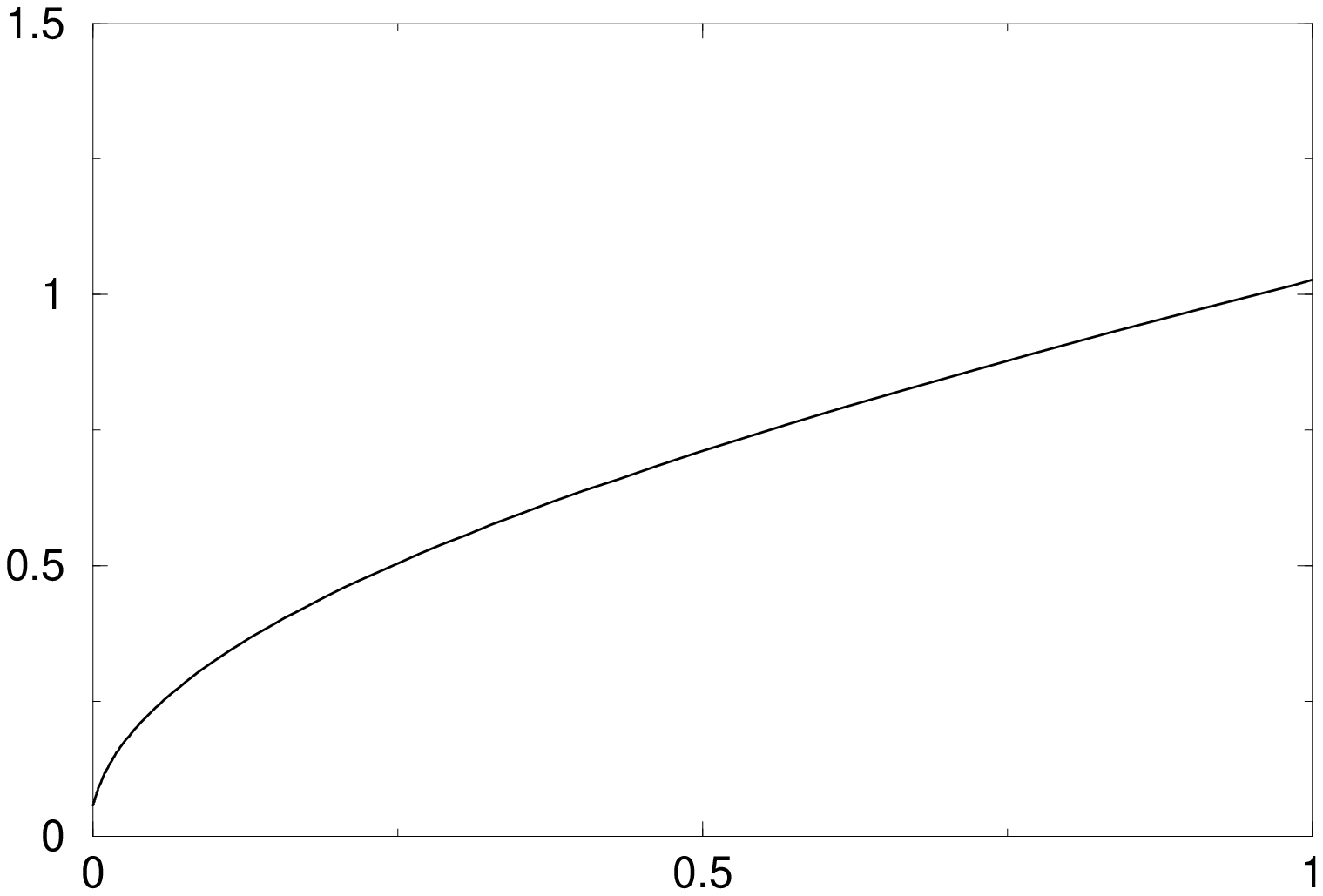,width=5cm,height=6cm}} 
\put(-40,-60){\epsfig{file=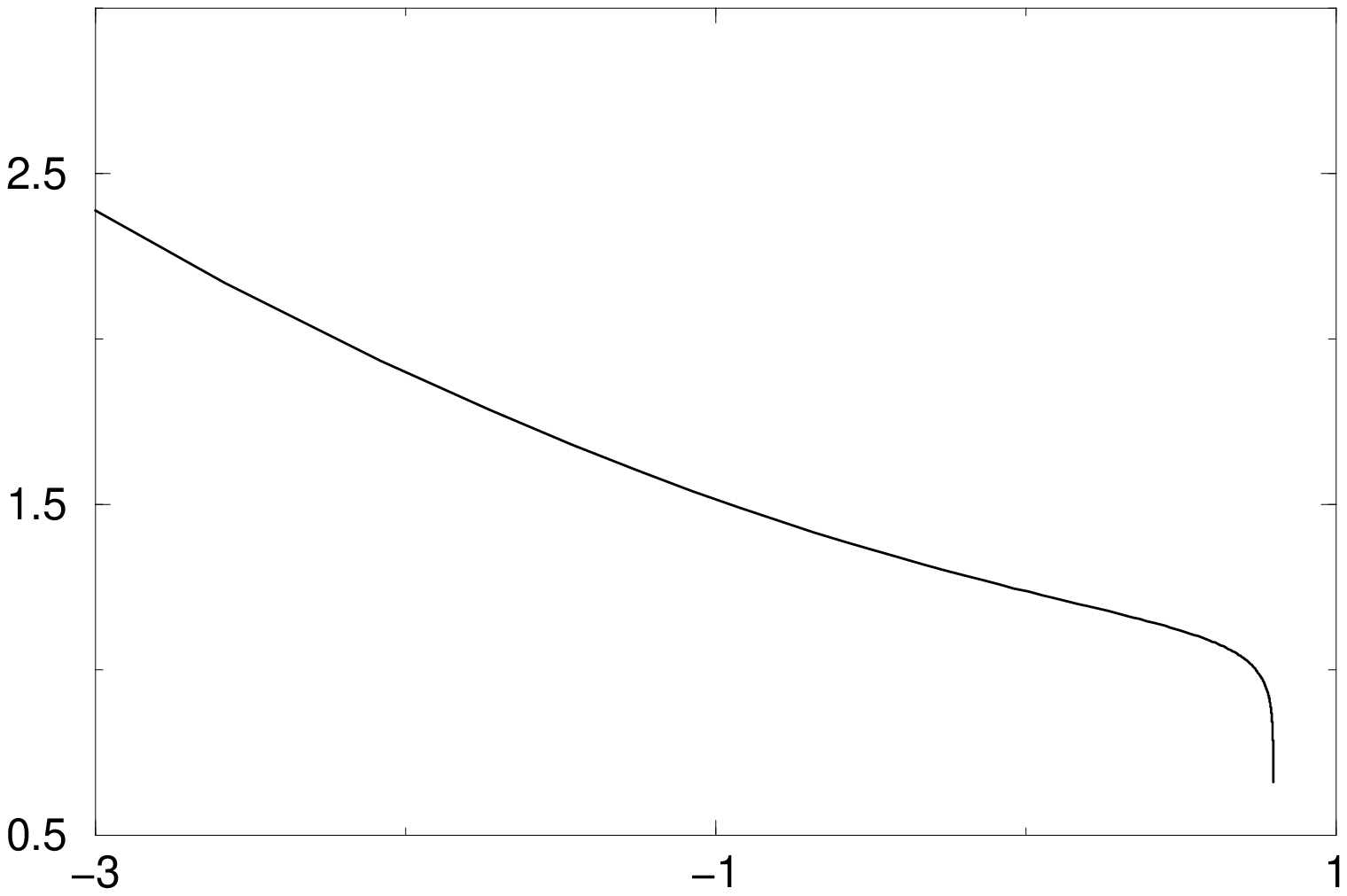,width=5cm,height=6cm}} 
\put(15,-60){\epsfig{file=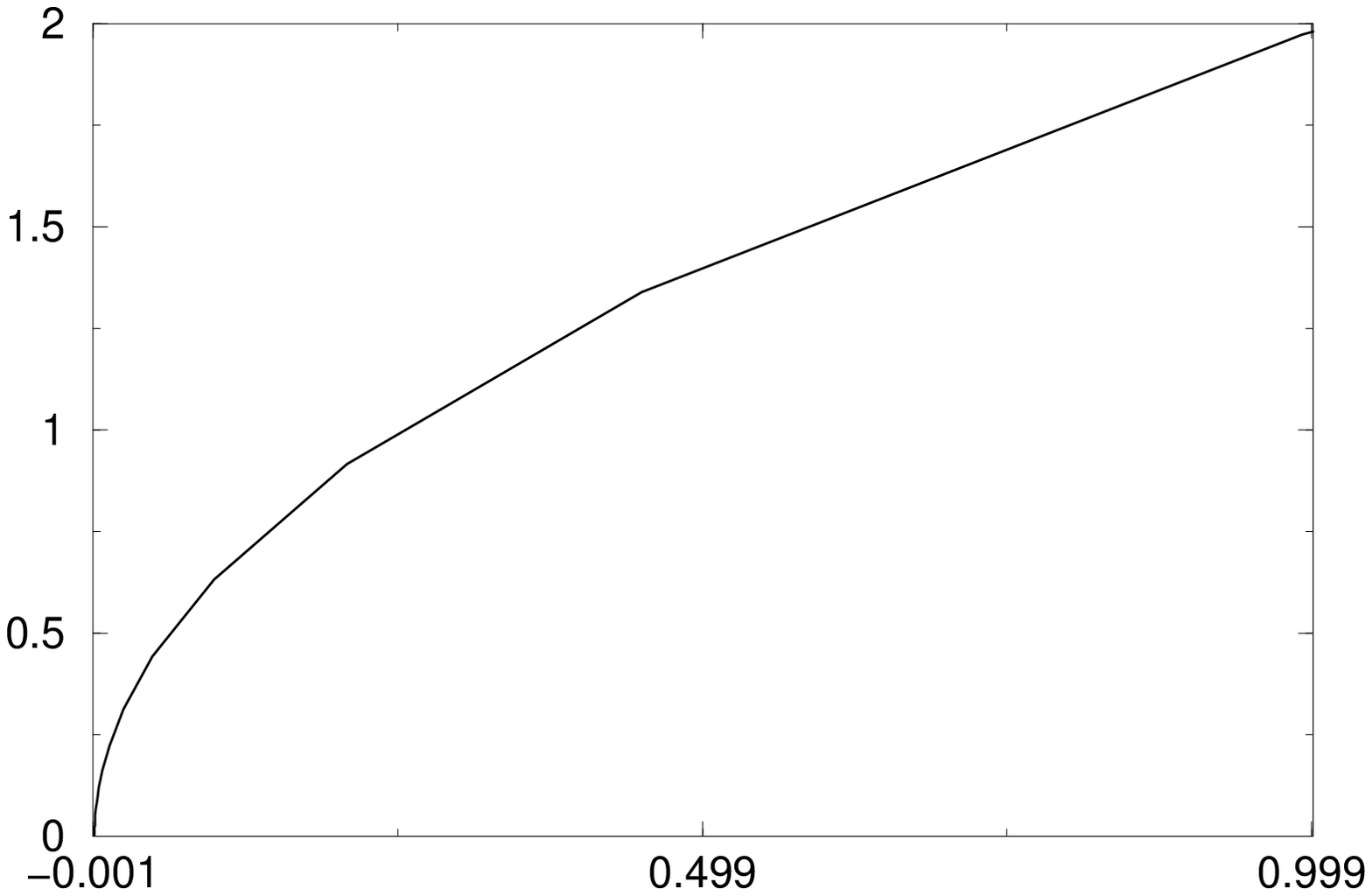,width=5cm,height=6cm}} 
\put(-35,5){$(a)$ phase $VIII-$} 
\put(20,5){$(b)$ phase $VIII+$} 
\put(75,5){$(c)$ phase $IX-$} 
\put(-35,-65){$(d)$ phase $IX+$} 
\put(20,-65){$(e)$ phase $X$} 
\end{picture} 
\end{center} 
\vspace{6cm} 
\caption{The behavior of the scale factor from phase $VIII$ to $X$} 
\end{figure}  
\vskip 0.5cm 

\section{Acceleration / Deceleration phase} 
Notice that as we have seen in Eq.(\ref{39}) not only the sign of
$H_{\pm}/T_{\pm}$ but also 
that of $H_{\pm}/T_{\pm} -1$ is important because the universe 
will accelerate or deccelerate according to the sign of the latter.
\subsection{$\go < 0 $ case} 
\subsubsection{$H_{-}/T_{-} > 1$}  
For $T_ - > 0$, the condition $H_{-}/T_{-} > 1$ is reduced to 
\be 
(3\gm+1)\sqrt{3(2\om+3)} < 3(1-\gm)(2\om+3) . \label{39}
\ee 
Consider  $\gm > -\frac{1}{3}$ case first. To our surprise, the inequality 
(\ref{39}) gives  us 
$\om > \frac{4(3\gm-1)}{3(1-\gm)^2}=\omega_{\Gamma_1}$, namely 
$\Gamma_1>0$. This is contradiction. 
Now for $\gm < -\frac{1}{3}$,
Eq.(\ref{39}) gives $\om < \om_{\go}$. The region I in figure
3 corresponds to this case. 

For $T_{-} < 0$, the  $H_{-}/T_{-} > 1$  is reduced to 
\be
(3\gm+1)\sqrt{3(2\om+3)} > 3(1-\gm)(2\om+3) .
\ee
This inequality gives $\om < \omega_{\Gamma_1}$ for $\gm > -\frac{1}{3}$.
The region II corresponds to this.
For $\gm < -1/3$,
we get the condition $\om > \om_{\go}$ which contradicts to $\go < 0$. 
In a summary, region I and II satisfies $H_- /T_- > 1$, $\go < 0$.
\subsubsection{ $H_{+}/T_{+} > 1$ }
For $T_+ > 0$, the condition $H_{+}/T_{+} > 1$ is reduced to  
\be 
(3\gm +1)\sqrt{3(2\om+3)} < -3(1-\gm)(2\om+3) .\label{41}
\ee 
The analysis is completely similar to the case A.
For $\gm < -1/3$, from above inequality we get $\om < \om_{\go}$ .
Since $T_+ > 0$ is satisfied only by the region II, III, IV, VII,
it is easy to see that there is no region satisfying all
three conditions, $\go < 0, T_+ > 0, \gm < -1/3$.
For $\gm > -1/3$ Eq.(\ref{41}) gives $\om > \om_{\go}$ which contradicts
to $\go¤< 0$.

For $T_{+} < 0$, $H_+ / T_+ > 1$ is reduced to
\be
(3\gm +1)\sqrt{3(2\om+3)} > -3(1-\gm)(2\om+3) .
\ee
When $\gm > -1/3$,
the above inequality gives $\om < \om_{\go}$. 
Since the condition $T_+ < 0$ is only satisfied by the region I, V, VI,
there is no region satisfying three conditions $\go < 0, T_+ < 0, \gm >
-1/3$.  
When $\gm < -1/3$, Eq.(\ref{41}) gives  
$\gm > \om_{\go}$ which contracdicts to $\go < 0$. 
In a summary, it is always $H_+ /T_+ < 1$, $\go < 0$.
\subsection{$\go > 0$ case} 
\subsubsection{$H_{-}/T_{-} >1$}  
For $T_{-} > 0$, the condition $H_- /T_- > 1$ is reduced to 
\be 
(3\gm +1) \sqrt{3(2\om+3)} < -3(1-\gm)(2\om+3) . \label{43} 
\ee 
Consider $\gm > -1/3$ case first.
The left hand side is always positive while the right hand side 
is always negative. So there is no solution for this.
For $\gm < -1/3$ case. Eq.(\ref{43}) gives $\om < \om_{\go}$
which contradicts to $\go > 0$.

For $T_- < 0$ case, from section III, we know that
there is no region satisfying $T_- < 0$ condition.
Therefore there are no solutions for conditions 
$\go > 0, H_- /T_- > 1$. 
We summarize if $\go > 0$ then we have $H_- /T_- < 1$ for
all region.   
\subsubsection{$H_{+}/T_{+} >1$}  
For $T_{+} > 0$, $H_+ /T_+ > 1$ is reduced to 
\be 
(3\gm+1)\sqrt{3(2\om+3)} < 3(1-\gm)(2\om+3) . \label{44}
\ee 
Consider $\gm > -1/3$ case. The above inequality gives
$\om > \om_{\go}$. Part of the region V 
satisfies these conditions.
For the solution to $\gm < -1/3$ case,
Eq.(\ref{44})  gives $\om < \om_{\go}$ which
contradicts to $\go > 0$.

For $T_{+} < 0$, $H_+ /T_+ >1$ is reduced to
\be
(3\gm+1)\sqrt{3(2\om+3)} > 3(1-\gm)(2\om+3) . \label{45}
\ee
When $\gm < -1/3$, we find solution $\om >
\om_{\go}$ which is satisfied by the region VI in Fig 3.
For $\gm > -1/3$ case, Eq.(\ref{45}) gives
$\om < \om_{\go}$ which contradicts $\go > 0$.
 
We summarize what we have obtained so far by  
a table. \\

\begin{tabular}{|c|c|c|c|c|c|c|}  
\hline 
phase & sign of & sign of & sign of & \hspace{0.5cm} range of 
\hspace{0.5cm} & 
$H_- /T_-$ & $H_+ /T_+$ \\  
  & $\go$ & $T_-$ & $T_+$ & t &  &   \\ \hline \hline 
I & $-$ & + & $-$ &$[t_i,t_f]$ & $H_- /T_- >1$ &$0<  H_+ /T_+ < 1$ \\
\hline 
II & $-$ & $-$ & + & $(-\infty,\infty)$ &$ H_- /T_- > 1$ & 
$0 < H_+ /T_+ < 1$ \\ \hline 
$III^-$ & + & + &  & $[t_i , t_f]$ & $0< H_- /T_- < 1$ & $ H_+ /T_+ >1$ \\ 
\hline 
$III^+$ &+ & & + & $[t_i,\infty]$ &  &   \\ \hline 
$IV^-$ &+ & + &  & $[t_i,t_f]$ & $ 0 < H_- /T_- < 1$ & $ H_+ /T_+ > 1$  \\ 
\hline 
$IV^+$ & + &  & + & $[t_i,\infty)$ &  &  \\ \hline 
$V^-$ & + & + &  & $[t_i,\infty)$ & $0 < H_- /T_- < 1$ & $ H_+ /T_+ > 1$ \\ 
\hline 
$V^+$ & + &  & + & $(-\infty,\infty)$ &  &  \\ \hline 
$VI^-$ & + & + &  & $[t_i,t_f]$ & $0 < H_- /T_- < 1$ &$H_+ /T_+ < 0$ \\ 
\hline 
$VI^+$ & + &  & $-$ & $[t_i,\infty)$ &  &  \\ \hline 
VII & $-$ & + & + & $[t_i,\infty)$ & $ H_- /T_- < 0$ &  
$0< H_+ /T_+ < 1$  \\ \hline 
$VIII^-$ & + & + &  & $[t_i,\infty)$ & $ 0 < H_- /T_- < 1$ &  
$ H_+ /T_+ < 0$  \\ \hline 
$VIII^+$ & + &  & $-$ & $(-\infty,t_f]$ &  &  \\ \hline  
$IX^-$ & + & + &  & $[t_i,\infty)$ & $0< H_- /T_- < 1$ & 
$ 0 < H_+ /T_+ < 1$  \\ \hline 
$IX^+$ & + &  & $-$ & $(-\infty,t_f]$ &  &  \\ \hline 
X & $-$ & + & + & $[t_i,\infty)$ & $ 0 < H_- /T_- < 1$ & 
$ 0 < H_+ /T_+ < 1$  \\ \hline 
\end{tabular} \\ 
 
\section{Einstein frame} 
In this section, we study the cosmology in Einstein frame. Especially, 
we investigate the difference between  the behavior of 
the scale factor in Einstein frame and that in the string frame 
as well as the possibity to avoid the initial singularity.    
 
The metric in Einstein frame is obtained from  
string frame metric by transformation using the relation $g_{E\mu\nu}   
= e^{-\phi}g_{\mu\nu}$:
\ba 
{ds_{E}}^2 &=&e^{-\phi}ds^2 \no 
&=& -e^{-\phi} dt^2 + e^{2\al-\phi}dx_i dx^i \no
 &=&-{dt_{E}}^2 + e^{2\al_E}dx_i dx^i \;\;(i=1,2,3) . \label{49}  
\ea 
From above relations, we see $\al_E = \al - \frac{\phi}{2}$ and  
$dt_E =e^{-\frac{\phi}{2}}dt$. 
Then we can obtain solutions in Einstein frame combining the above relation  
with the solutions in string frame. 
Therefore the solutions in Einstein frame are 
\be 
\al_E(\tau) =c \frac{(3\gm+1)\sqrt{3(2\om+3)}}{-4\go}\tau - \ln \left[
\frac{q}{c} 
\cosh(c\tau) \right] \left( \frac{(3\gm+1)\gt +4\go}{8\go} \right) + 
\frac{3\gm+1}{4}B, \label{al1}  
\ee 
for $\go < 0$, and  
\be 
\al_E(\tau) =c \frac{(3\gm+1)\sqrt{3(2\om+3)}}{4\go}\tau - \ln \left[
\frac{q}{c} 
|\sinh(c\tau)| \right] \left( \frac{(3\gm+1)\gt +4\go}{8\go} \right) + 
\frac{3\gm+1}{4}B,  \label{al2}  
\ee 
for $\go > 0$. 
In next section we find out $t_E(\tau)$ and see that the interval
of $t_E(\tau)$ can be classified by $\om$ and $\gm$.
\subsection{Classification of the phases by $t_E$ and $\tau$} 
\subsubsection{$\go < 0$ case}  
From Eq.(\ref{a}) in section II, and using Eqs.(\ref{49}) and (\ref{al1}),  
$t_E(\tau)$ can be read 
\ba 
t_E -t_{E0} &=& \int d\tau e^{3\al_E(\tau)} \no 
& \sim & \int d\tau e^{\frac{3(3\gm+1)\sqrt{3(2\om+3)}}{-4\go}c\tau} 
\left[ \cosh(c\tau) \right]^{-\frac{3[4\go +(3\gm+ 1)\gt]}{8\go}} .
\ea 
In the limit $\tau \to \pm \infty$, we can write 
$$t_E \sim \frac{1}{T_{\pm}} e^{T_{E\pm}\tau}.$$ 
where  
\be 
T_{E\pm} = -\frac{3(3\gm+1)\sqrt{3(2\om+3)}}{4\go} \mp  
\frac{3[4\go +(3\gm+1)\gt]}{8\go} .
\ee 

The condition for $T_{E-} < 0$ is reduced to
\be
(3\gm +1)\sqrt{3(2\om+3)} < 3(1-\gm)(2\om+3). \label{t1}
\ee
Consider first $\gm > -\frac{1}{3}$ case. The solution for Eq.(\ref{t1}) 
is $\om > \om_{go}$ which violate $\go < 0$. For $\gm < -\frac{1}{3}$,
we get $\om < \om_{\go}$.
Now $T_{E-} > 0$ case that is
\be
(3\gm+1)\sqrt{3(2\om+3)} > 3(1-\gm)(2\om+3). \label{t2}
\ee
For $\gm > -\frac{1}{3}$, Eq.(\ref{t2}) gives $\om < \om_{\go}$.
However, for $\gm < -\frac{1}{3}$, the left hand side
is negative while right hand side is positive which is inconsistent.   

Now consider $T_{E+} < 0$ case which is reduced to
\be
(3\gm+1)\sqrt{3(2\om+3)} < -3(1-\gm)(2\om+3). \label{t3}
\ee
For $\gm > -\frac{1}{3}$, in Eq.(\ref{t3}) the
left hand side is positive while right hand side negative which is
inconsitent.
For $\gm < -\frac{1}{3}$, we have $\om < \om_{\go}$.
Consider $T_{E+}$ which is reduced to
\be
(3\gm+1)\sqrt{3(2\om+3)} > -3(1-\gm)(2\om+3). \label{t4}
\ee
For $\gm > -\frac{1}{3}$, we have $\om < \om_{\go}$. For
$\gm < -\frac{1}{3}$, Eq.(\ref{t4}) gives $\om > \om_{\go}$ which
violates $\go < 0$.   
One can summarize that under the condition $\om < \om_{\go}$,
$$ {\rm For} \;\; \gm < -\frac{1}{3}: \;\; T_{E-} < 0,\;\; T_{E+} < 0.$$
$$ {\rm For} \;\; \gm > -\frac{1}{3}: \;\; T_{E-} > 0,\;\; T_{E+} > 0.$$  
These relations tell us that
$$ {\rm For} \;\; \gm < -\frac{1}{3}: \;\; -\infty < t_E < t_{Ef}.$$
$$ {\rm For} \;\; \gm > -\frac{1}{3}: \;\; t_{Ei} < t_E < \infty.$$
We emphasize that there is no region where $t_E$ can run from $-\infty$
to $+\infty$.
This is because $T_{E+}$ and $T_{E-}$ have the same sign in any given
region unlike the string frame.
\subsubsection{$\go > 0$ case}  
In this case $t_E(\tau)$ can be read from Eqs.(\ref{a}), (\ref{49}) and
(\ref{al2}),   
\ba 
t_E -t_{E0} &=& \int d\tau e^{3\al_E(\tau)} \no 
& \sim & \int d\tau e^{\frac{3(3\gm+1)\sqrt{3(2\om+3)}}{4\go}c\tau} 
|\sinh(c\tau)|^{-\frac{3(4\go +(3\gm+ 1)\gt)}{8\go}} . 
\ea 
As we saw above, this case is singular as $\tau \to 0$. So it  
is necessary to consider the behavior in that case. As $\tau \to 0$, 
\be 
t_E  \sim {\rm sign}(\tau)\frac{|\tau|^{1-\eta} }{1-\eta}, \label{61} 
\ee 
where $\eta = \frac{3[(3\gm+1)\gt + 4\go]}{8\go}$. 
If $\eta > 1$, $t_E$ is singular at $\tau =0$, 
while if $\eta < 1$, it is regular. 
$\eta > 1$ case gives 
$$\om > -\frac{(5+3\gm)}{3(1-\gm^2)}=\om_\star.$$
The other case, $\eta < 1$, 
gives $\om < \om_\star$  which does not overlap 
with $\go > 0$. 
Therefore there is no regular region for $\go > 0$.  
As $\tau \to -\infty$, $t_E \to +\infty$ while
as $\tau \to +0$, $t_E \to -\infty$.
Let us find out the behavior in the region $\tau \to \pm  
\infty$. In this limit $t_E$ and $\tau$ is given by 
\be 
t_E - t_{E0} \sim  \int d\tau e^{T_{E\pm}\tau},
\ee 
where 
\be 
T_{E\pm} = \frac{3(3\gm+1)\sqrt{3(2\om+3)}}{4\go} \mp 
\frac{3[(3\gm+1)\gt + 4\go]}{8\go} . 
\ee 

The condition $T_{E-} > 0$ is reduced to
\be
(3\gm+1)\sqrt{3(2\om+3)} > -3(1-\gm)(2\om+3) .\label{ll}
\ee
For $\gm > -\frac{1}{3}$, the left hand side is always positive 
while the right hand side always negative. So Eq.(\ref{ll}) satisfies
trivially.
For $\gm < -\frac{1}{3}$, Eq.(\ref{ll}) gives $\om > \om_{\go}$
which is consistent with $\go > 0$.  

For $T_{E-} < 0$, we have the inequality
\be
(3\gm+1)\sqrt{3(2\om+3)} < -3(1-\gm)(2\om+3) . \label{lm}
\ee
For $\gm > -\frac{1}{3}$, since after dividing by $3\gm+1$ the left hand
side is
always positive while the right hand side is always negative.
So there is no solution. For $\gm < -\frac{1}{3}$, Eq.(\ref{lm})
gives $\om < \om_{\go}$ which contradicts to $\go > 0$.

Now consider the case $T_{E+} > 0$.
This condition is reduced to
\be
(3\gm+1)\sqrt{3(2\om+3)} > 3(1-\gm)(2\om+3) . \label{ln}
\ee
Consider first $\gm > -\frac{1}{3}$. Eq.(\ref{ln}) gives
$\om < \om_{\go}$ which contradicts to $\go > 0$. For
$\gm < -\frac{1}{3}$, the left hand side is always positive while the
right hand side always
is negative. So we have no solution for $T_{E+} > 0$.

Consider $T_{E+} < 0$ which is reduced to
\be
(3\gm +1)\sqrt{3(2\om+3)} < 3(1-\gm)(2\om+3) . \label{lo}
\ee
For $\gm > -\frac{1}{3}$, Eq.(\ref{lo}) gives $\om > \om_{\go}$
which satisfies $\go > 0$. For $\gm < -\frac{1}{3}$, in Eq.(\ref{lo})
after dividing by $3\gm+1$ we see that the left hand side  is positive
while the right hand side is negative. 
It is easy to check that when $\om > \om_{\go}$,
$$ T_{E-} < 0, \;\; and \;\; T_{E+} < 0.$$  
In a summary, 
for $\go > 0$, $t_E$ is singular as $\tau \to  0$. 
$t_E$ runs from initial time $t_{Ei}$ to $+\infty$ for 
$\tau \in (-\infty , 0)$ and $-\infty$ to final time
$t_{Ef}$ for $\tau \in (0, \infty)$. 
\subsection{The scale factor}   
In previous section we see that $t_E(\tau)$ is not 
monotonic function which is crucial to decide whether the scale
factor is singular or not. In this section following the same procedure
we study the behavior of the 
scale factor, $a_E(\tau)$, and classify by $\om$ and $\gm$.  
 
\subsubsection{$\go < 0$ case} 
For this case we use the Eq.(\ref{al1}), then 
\ba 
a_E(\tau)&=&e^{\al_E(\tau)} \no 
&=&c_1 e^{c \frac{(3\gm+1)\sqrt{3(2\om+3)}}{-4\go}\tau} 
\left[ \frac{q}{c} \cosh(c\tau) \right]^{-\frac{(3\gm+1)\gt +4\go}{8\go}}, 
\ea 
where $c_1= e^{\frac{(3\gm+1)B}{4}}$.

In the limit $\tau \to \pm \infty$, the scale factor can be  
rewritten as $e^{H_{E\pm}\tau}$ where  
\be 
H_{E\pm} = -\frac{(3\gm +1) \sqrt{3(2\om+3)}}{4\go}  
\mp \frac{(3\gm +1)\gt+4\go}{8\go} 
\ee 
Since, in the limit $\tau \rightarrow \pm \infty$,   
$H_E$ and $T_E$ are proportional$(T_E = 3H_E)$, we can analyse by 
using $T_E$ in the previous subsection. 

Summary:
$$a_E(\tau) \;\;runs\;\; from \;\;+\infty \;\;to\;\; 0\;\; for\;\;
 \gm < -\frac{1}{3}.$$ 
$$a_E(\tau) \;\;runs\;\; from\;\; 0 \;\;to\;\; +\infty\;\; 
for \gm > -\frac{1}{3}.$$
 
\subsubsection{$\go > 0$ case}  
From Eq.(\ref{al2}), we can write the scale factor for this case as follows.  
\ba 
a_{E}(\tau) &=& e^{\al_E(\tau)} \no  
&=& c_2 e^{\frac{(3\gm+1)\sqrt{3(2\om+3)}}{4\go}c\tau}
|\sinh(c\tau)|^{-\frac{[(3\gm+1)\gt +4\go]}{8\go}} . 
\ea
where $c_2= e^{\frac{(3\gm+1)B}{4}} 
(\frac{q}{c})^{-\frac{[(3\gm+1)\gt+4\go]}{8\go}} $.
Since as we saw in the section VII.1.B, $\eta < 1$ and $\go > 0$
are not consitent with each other, we consider the case $\eta > 1$.
In this region there is singularity at $\tau =0$. So 
we need to consider the limit $\tau\to 0$: 
\ba 
a_E(\tau) &\to& |\tau|^{-\eta /3} \no 
&=& |\tau|^{-\frac{(3\gm+1)\gt+4\go}{8\go}}. \label{71}  
\ea 
As $\tau \to 0$, the behavior of the
scale factor always goes to positive infnity because $\eta > 1$.   
As $\tau \rightarrow \pm \infty$, 
$a_{E}(\tau)$ can be written as 
$$e^{H_{E\pm}\tau}$$ 
where 
\be 
H_{E\pm} = \frac{(3\gm+1)\sqrt{3(2\om+3)}}{4\go} \mp 
\frac{[(3\gm+1)\gt+4\go]}{8\go}. 
\ee 
By the same analysis of subsection VII.1.B  we can write the solutions.
Under the condition $\om > \om_{\go}$, 
$$H_{E-} > 0 \;\; and \;\; H_{E+} < 0.$$  
The scale factor evolves from 0 to $+\infty$ 
for $\tau \in  (-\infty, 0)$ and from $\infty$ to  
0 for $\tau \in (0,\infty)$. 

Now we study $a_E(t_E)$.
First consider in the limit $\tau \to 0$. In this limit $\go < 0$ case
is regular. Therefore we investigate $\go > 0$ case. From Eqs.(\ref{61})
and (\ref{71}) we write $a_E(t_E)$ as
\be
a_E(t_E) \sim \left[ (1-\eta)sign(\tau) t_E \right]^{\frac{\eta}{3(\eta -1)}}.
\ee
When $\frac{\eta}{3(\eta-1)} > 1$ i.e. $1 < \eta < \frac{3}{2}$, the 
scale factor will accelerate while when $\frac{\eta}{3(1-\eta)} < 1$ i.e. 
$\eta > \frac{3}{2}$ the scale factor will decelerate.
To accelerate, $\om$ should satisfy
$$\om > \om_{\gt} \;\; for \;\; \gm < -\frac{1}{3}.$$
$$\om < \om_{\gt} \;\; for \;\; \gm > -\frac{1}{3}.$$
In Fig.7, region IV correspond to these relations.
For deceleration $\om$ should satisfy
$$\om > \om_{\gt} \;\; for \;\;\gm > -\frac{1}{3}.$$
$$\om < \om_{\gt} \;\; for \;\; \gm < -\frac{1}{3}.$$
Region III and V satisfy these relations.

In both cases($\go >0 \;\; or \;\; \go <0$), in the limit 
$\tau \to \pm \infty$ the scale factor $a_E(t_E)$ behaves as
\be
a_E(t_E) \sim {t_E}^{H_{E\pm}/T_{E\pm}} . \label{65}
\ee
Since we already know that $T_E$ and $H_E$ satisfy $T_{E\pm} = 3H_{E\pm}$,
from Eq.(\ref{65}), $a_E(t_E)$ can be written 
$$ a_E(t_E) \sim {t_E}^{1/3} .$$
It is interesting to compare with Einstein general relativity 
where $a_E(t_E)$ behaves as
$$a_E(t_E) \sim {t_E}^{2/3} \;\; for \;\; dust \;\;(\gm=0).$$
$$a_E(t_E) \sim {t_E}^{1/2} \;\; for \;\; radiation \;\; (\gm=1/3).$$
 
\begin{figure}[hbt]
\centerline{\epsfig{file=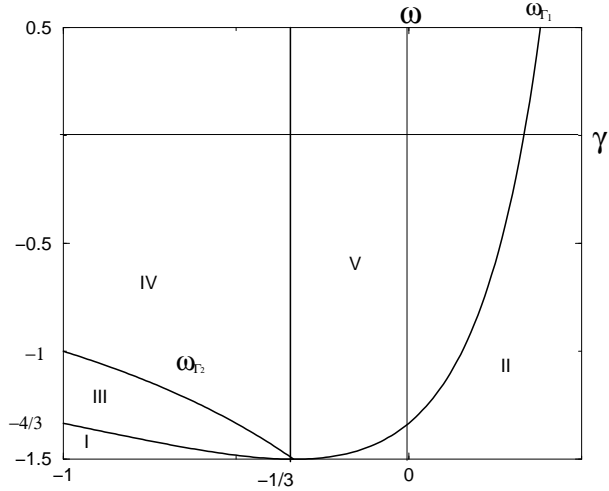,width=8cm}}
\caption{\small The classification of phase space in Einstein frame}
\end{figure}
 
We summarize the behavior of the scale factor in terms of $t_E$.  
\begin{itemize}
\item Region I: $T_{E-}(=3H_{E-}) < 0, T_{E+}(=3H_{E+}) < 0$.  
The universe described by $a_E(t_E)$ evolves from infinite size to zero 
as time runs from negative infinity to finite final time $t_{Ef}$.  
\item Region II: $T_{E-} > 0, T_{E+} > 0$. 
The universe evolves from zero size to 
infinite one as time runs from finite initial time $t_{Ei}$ 
to positive infinity. 
\item Region III, IV and V: $T_{E-} > 0$ and $T_{E+} < 0$. The  
universe evolves from zero to infite size as time runs 
from finite initial time $t_{Ei}$ to infinity for $\tau \in (-\infty, 0)$
and infinte size to
zero as time runs from negative infinity to finite 
final time $t_{Ef}$ for $\tau \in (0, +\infty)$. 
Furthermore, in the limit $\tau \to
0$ the universe has acceleration or deceleration regime
which depends on the range of $\eta$. Region
III and V is decelerationary  while region IV is accelerationary
phase.
 
\end{itemize}    
 
\section{Discussion and conclusion}  
We have considered string motivated Brans-Dicke(BD) cosmology 
with perfect fluid type matter which arise when a certain kind of 
the dilaton coupled p-brane gas is dominating the universe.
This is the complementary study to our earlier
work\cite{sung,park}, where p-brane gas that does not couple 
to the dilaton was studied. 
Cosmology is classified into 16 phases according to the asymptotic
behavior of the time interval and the scale factor.
This is qualitatively similar
to the result obtained before for the dilaton coupled
case\cite{sung,park}.
In string frame, there is a phase where the cosmology has no singularity,
namely, region II in Fig 3.
In Einstein frame, contrary to the string frame, there is 
no singularity free phase. This is partly due to the difference of cosmic
time($t$ in string frame and $t_E$ in Einstein frame) and
partly due to the dilaton factor relating two frames.
In asymptotic regime, $\tau \to \pm \infty$, the behavior
of the scale factor is ${t_E}^{1/3}$.
To our surprise the inflationary regime of the
dilaton-graviton string cosmology is gone in
the presence of the matter. The matter
contribution seems to give mass term or potential to the dilaton
regulating the dilaton from growing dilaton kinetic energy\cite{ve}. 

In a recent study\cite{bak}, with the assumption of holographic principle, 
it was argued that this principle requires the existence of graceful exit 
by smoothly connecting the pre and post big-bang branches.   
According to the \cite{park1}, all cosmological solutions of
p-brane dominating the universe can be mapped to
the present case or the case studied in \cite{sung,park}.
In any case, there is no solution which exhibit both inflation and
graceful exit. Therefore our result draws a negative conclusion to what
has been said in \cite{bak}.

\end{document}